\documentclass{article}
\usepackage{epsfig,amssymb}
\input{epsf}
\newcommand{\be}{\begin{equation}}
\newcommand{\ee}{\end{equation}}
\newcommand{\ba}{\begin{eqnarray}}
\newcommand{\ea}{\end{eqnarray}}
\newcommand{\beq}{\begin{equation}}
\newcommand{\eeq}{\end{equation}}
\newcommand{\beqa}{\begin{eqnarray}}
\newcommand{\eeqa}{\end{eqnarray}}

\newcommand{\vs}{\vspace{-0.15cm}}
\begin{document}

\begin{flushright}
{\tiny  FZJ-IKP(TH)-2001-05} \\
{\tiny  UK/TP-2001-13} \\
\end{flushright}

\vspace{1in}

\begin{center}

\bigskip

{{\Large\bf Rescattering and chiral dynamics in
    \boldmath{$ B \to \rho\pi$} decay    }}

\end{center}

\vspace{.3in}

\begin{center}
{\large
S. Gardner$^{\dagger,}$\footnote{email: gardner@pa.uky.edu} and
Ulf-G. Mei{\ss}ner$^{\ddagger,}$\footnote{email: u.meissner@fz-juelich.de}}

\vspace{1cm}

$^\dagger${\it Department of Physics and Astronomy, University of Kentucky\\
Lexington, Kentucky 40506-0055, USA}\\

\bigskip

$^\ddagger${\it Forschungszentrum J\"ulich, Institut f\"ur Kernphysik
(Theorie)\\ D-52425 J\"ulich, Germany}

\bigskip

\end{center}

\vspace{.6in}

\thispagestyle{empty}

\begin{abstract}\noindent
We examine the role of
$B^0(\bar B^0) \to \sigma \pi^0 \to \pi^+\pi^- \pi^0$ decay in the
Dalitz plot analysis of $B^0 (\bar B^0) \to \rho\pi \to \pi^+\pi^-\pi^0$
decays, employed to extract the CKM parameter $\alpha$. 
The $\sigma \pi$ channel is significant because it
can break the relationship between the penguin contributions
in $B\to\rho^0\pi^0$, $B\to\rho^+\pi^-$, and $B\to\rho^-\pi^+$
decays consequent
to an assumption of isospin symmetry. Its presence thus
mimics the effect of isospin violation.
The $\sigma\pi^0$ state is of definite CP, however; we demonstrate
that the $B\to\rho\pi$ analysis can be generalized
to include this channel without difficulty.
The $\sigma$ or $f_0(400-1200)$ ``meson'' is a
broad $I=J=0$ enhancement driven by strong $\pi\pi$ rescattering;
a suitable scalar form factor is 
constrained by the chiral dynamics of low-energy
hadron-hadron interactions --- it is rather different from
the relativistic Breit-Wigner form adopted in earlier $B\to\sigma\pi$ and
$D\to\sigma\pi$ analyses. We show that the use of this
scalar form factor leads to an improved theoretical
understanding of the measured ratio
${\rm Br}({\bar B^0} \to \rho^\mp \pi^\pm ) /
{\rm Br}({B^-} \to \rho^0 \pi^- )$.
\end{abstract}

\vfill

{\noindent{\it Accepted for publication in Physical Review D.}}

\pagebreak

\section{Introduction}
Measurements at SLAC and KEK of the
time-dependent CP-violating
asymmetry in $B(\bar B) \to J/\psi K_s$~\cite{babar,belle},
yielding $\sin(2\beta)$,
have conclusively established the existence of CP violation in
the B meson system. The results found are consistent with
Standard Model (SM) expectations~\cite{buras}, so that
establishing whether the Cabibbo-Kobayashi-Maskawa (CKM)
matrix~\cite{ckmmatrix} is the
only source of CP violation in nature, as in the SM,
or not requires the empirical measurement of all
the angles of the unitarity triangle.

In this paper we consider the determination of $\alpha$
through a  Dalitz plot analysis of the decays
$B^0 ({\bar B}^0) \to \rho \pi \to \pi^+ \pi^- \pi^0$ under the
assumption of isospin symmetry~\cite{LNQS,QS1}.
Ten parameters appear in the analysis, and they
can be determined in a fit to the data. Nevertheless,
the assumption of $\rho$ dominance in $B\to 3 \pi$ decays
has no strong theoretical basis~\cite{babarbook},
so that the contributions from other
resonances in the $\rho \pi$  phase space may be important.
We discuss how the isospin analysis can be
enlarged to include the $\sigma\pi$ channel as well.
The $\sigma$  or $f_0(400-1200)$ ``meson'' is a 
broad $J=I=0$ enhancement, close to the $\rho$ meson
in mass, so that the $\sigma\pi$ channel can potentially populate
the $3\pi$ phase space associated with the $\rho\pi$ channels.
The $\sigma\pi$ final state contributes preferentially
to the $\rho^0\pi^0$ final state.
In the context of the isospin analysis, such contributions are
of consequence as they invalidate
the underlying assumptions of the isospin
analysis and thus mimic the effect of isospin violation.

Our considerations are inspired in part by recent
studies of $D^-\to \pi^- \pi^+\pi^-$ decay:
the E791 collaboration find that the pathway
$D^-\to \pi^- \sigma \to \pi^- \pi^+ \pi^-$ accounts
for approximately half of all
$D^-\to \pi^- \pi^+\pi^-$ decays~\cite{e791ex}.
Deandrea and Polosa have
argued as a consequence that the $B\to \sigma\pi$ channel contributes
significantly to the $\rho\pi$ phase space in $B\to \pi \pi^+\pi^-$
and modifies the ratio
${\cal B}(\bar{B^0} \to \rho^\mp\pi^\pm)/{\cal B}(B^- \to \rho^0\pi^-)$
to yield better agreement with experiment~\cite{deA}.
The scalar form factor, which describes the appearance of the
$\sigma$ in the $\pi^+\pi^-$ final state, enters as a
crucial ingredient in the assessment of the size of these effects.
The scalar form factor cannot be determined directly
from experiment; nevertheless,
ample indirect constraints exist, permitting us to describe
its features with confidence~\cite{OM}.
Nevertheless, different approaches, with different dynamical
assumptions, yield roughly comparable descriptions of the
$\pi\pi$ scattering data,
so that the emergence of a favored form of the scalar form
factor does not resolve the question of whether the $\sigma$
is a pre-existing resonance or, rather, a dynamical
consequence of $\pi\pi$ interactions in the final-state.
We follow Ref.~\cite{OM} and
adopt an unitarized, coupled-channel approach to the final-state
interactions (FSI) in the $\pi\pi$-$K\bar K$ system, and match the resulting
scalar form factor to chiral perturbation theory (CHPT)
in the regime where the latter is applicable.
The resulting form factor, in the $\pi\pi$ channel, discussed
in Sec. 5, is strikingly
different from the relativistic Breit-Wigner form adopted by
the E791 collaboration in their analysis of the 
$\sigma$ in $D^+\to \pi^+ \pi^+ \pi^-$ decay --- the latter
form factor is also used in Ref.~\cite{deA}.
The differences are particularly large as $\sqrt{s}\to 2 M_\pi$,
so that the relativistic Breit-Wigner form is at odds with
CHPT in the precise region where it is
applicable, note Fig. 4 in Ref.~\cite{OM}. This casts doubt
on the recent conclusions of Refs.~\cite{e791ex,deA}, prompting
new analyses incorporating a suitable scalar form factor.

The generation of the $\sigma$ resonance
via strong rescattering effects,
as in the approach we adopt, indicates that
OZI-violating effects in the scalar sector
are significant. Moreover,
the ``doubly'' OZI-violating form factor
$\langle 0 | \bar s s | \pi \pi\rangle$ is non-trivial as well;
such a contribution is
needed to fit the $\pi\pi$ and $K\bar K$ invariant mass
distributions in $J/\psi \to \phi \pi\pi (K\bar K)$ decay~\cite{OM}.
These effects are also needed to explain the branching ratios of
the decays of the $a_0(980)$ and $f_0(980)$ states into $\pi\pi$ and
$KK$ final states~\cite{ollerrev}.
These observations give new insight on rescattering effects in
hadronic B decays, generating a new mechanism of factorization
breaking in $n\ge 3$ particle final states.

The contribution of the $B\to \sigma\pi$ channel to the $B\to \rho^0 \pi$
phase space can also modify the inferred empirical branching ratios
in these channels. Combining the CLEO results~\cite{Jessop:2000bv}
\begin{eqnarray}
\label{cleobr}
{\rm Br}({B^-} \to \rho^0 \pi^- ) &=&
(10.4^{\,+3.3}_{\,-3.4} \pm 2.1)\cdot 10^{-6} \;, \\
{\rm Br}({\bar B^0} \to \rho^\pm \pi^\mp ) &=&
(27.6^{\,+8.4}_{\,-7.4} \pm 4.2)\cdot 10^{-6}
\end{eqnarray}
with the BaBar result~\cite{Aubert:2001xb}
${\rm Br}({B^0} \to \rho^\pm \pi^\mp)=28.9 \pm 5.4\pm 4.3$
(charge conjugate modes are implied) 
yields, 
adding the errors in quadrature and ignoring correlations,
\begin{equation}
\label{rhoratio}
{\cal R} = \frac{{\rm Br}({\bar B^0} \to \rho^\mp \pi^\pm )}
{{\rm Br}({B^-} \to \rho^0 \pi^- )} = 2.7 \pm 1.2\;.
\end{equation}
This ratio of ratios is roughly 6 if one works at tree level
and uses the naive factorization approximation for the
hadronic matrix elements~\cite{bsw}.
The inclusion of penguin contributions can alter this
result, and potentially yield
better accord with theory and experiment~\cite{Ali98,dearho,hycheng,rhopqcd}.
However, our focus will parallel that of Ref.~\cite{deA}: we wish
to examine how $B\to\sigma\pi \to 3\pi$ decay, given a particular
scalar form factor, can effectively modify
the theoretical prediction of the ratio given in Eq.~(\ref{rhoratio}).
It is apparent that $B\to\sigma\pi$ is of greater impact in
$B\to\rho^0\pi$ decay, so that the inclusion of such contributions
ought alter the ratio of ratios.

We begin by reviewing the isospin analysis in
$B^0 (\bar B^0) \to\rho\pi \to \pi^+\pi^-\pi^0$
decay~\cite{LNQS,QS1} in Sec.~\ref{sec:isoanal}, and
discuss its extension to include
$B^0 (\bar B^0) \to\sigma\pi\to \pi^+\pi^-\pi^0)$
decay in Sec.~\ref{sec:extend}.
We proceed by evaluating $\sigma$-mediated $B\to 3\pi$ decay in
Sec.~\ref{sec:evalB3pi}, relegating the $\rho$-mediated
$B\to 3\pi$ decay formulae to App.\ref{app:Brhopi}. Our analysis
employs the scalar and vector form factors discussed in
Sec.~\ref{sec:sff} and Sec.~\ref{sec:vector}, respectively.
We conclude with a presentation of our
results in Sec.~\ref{sec:res} and an accompanying summary.

\section{Preliminaries: Isospin analysis of \boldmath{$B\to \rho\pi$}}
\label{sec:isoanal}
Let us recall the isospin analysis possible
in $B\to \rho\pi$ decay~\cite{LNQS,QS1}.
Under the assumption of isospin symmetry, a $\rho\pi$ final state
can have isospin $I_f=0,1$, or $2$, whereas the
$B^+, B^0$ states form an isospin doublet.
Thus we can have $|\Delta I|=1/2,3/2$, or $5/2$ transitions
in $B\to \rho\pi$ decay, so that we can parametrize the
amplitudes which appear by $A_{|\Delta I|,I_f}$.
We have~\cite{LNQS,QS}\footnote{We flip the overall
sign of the $a_{00}$ amplitude with respect
to Ref.~\cite{QS}, to conform with our computation of the amplitudes.}

\begin{equation}
\label{isopm}
a_{+-} \equiv A(B^0\to \rho^+\pi^-) =
\frac{1}{2\sqrt{3}}[ A_{3/2,2} + A_{5/2,2}] +
\frac{1}{2}[A_{3/2,1} + A_{1/2,1}] +
\frac{1}{\sqrt{6}} A_{1/2,0}\;,
\end{equation}
\begin{equation}
\label{isomp}
a_{-+} \equiv A(B^0\to \rho^-\pi^+) =
\frac{1}{2\sqrt{3}}[ A_{3/2,2} + A_{5/2,2}] -
\frac{1}{2} [A_{3/2,1} + A_{1/2,1}] +
\frac{1}{\sqrt{6}} A_{1/2,0}\;,
\end{equation}
and
\begin{equation}
\label{iso00}
a_{00} \equiv A(B^0\to \rho^0\pi^0) =
-\frac{1}{\sqrt{3}}[ A_{3/2,2} + A_{5/2,2}] +
\frac{1}{\sqrt{6}} A_{1/2,0}\;,
\end{equation}
noting that
$A(B^0\to \pi^+\pi^-\pi^0)=f_+\, a_{+-} + f_-\, a_{-+} + f_0\, a_{00}$,
where $f_i$ is the 
form factor describing $\rho^i \to \pi \pi$.
Isospin is merely an approximate symmetry
of the SM;
nevertheless, our parametrization
possesses three independent ``isospin'' amplitudes, distinguished by
$I_f$,
to describe the three empirical amplitudes $a_{ij}$, so that it
persists in the presence of isospin breaking as well.

The lowest-dimension operators of the 
effective, $|\Delta B|=1$ Hamiltonian
generate transitions of $|\Delta I|=1/2$ or $3/2$
character, so that a $|\Delta I|=5/2$
transition is generated, in this order,
through long-distance, isospin-breaking effects in concert with
a $|\Delta I|=1/2$ or $3/2$, short-distance, weak transition.
If we neglect transitions of $|\Delta I|=5/2$ character and,
indeed, isospin-violating effects all together,
the transition $\bar b\to q\bar q \bar d$ which mediates
$B\to \rho\pi$ decay can be realized through a
``tree'' amplitude with $|\Delta I|=1/2$ or $3/2$ or through
a ``penguin'' amplitude with $|\Delta I|=1/2$. Practically,
the decay topologies
are distinguished by their weak phase, so that the contributions
associated with the CKM factors
$V_{ub}^\ast V_{ud}^{}$, e.g., are defined
to be tree contributions, regardless of their dynamical origin.
The unitarity of the CKM matrix in the SM
implies that two combinations
of CKM factors suffice in describing $b\to q \bar q q^\prime$;
here we associate the combination $V_{tb}^\ast V_{td}^{}$ with the
penguin contribution. Noting
\begin{equation}
\frac{V_{ub}^\ast V_{ud}^{}}{|V_{ub}^\ast V_{ud}^{}|} = e^{i\gamma} \quad;\quad
\frac{V_{tb}^\ast V_{td}^{}}{|V_{tb}^\ast V_{td}^{}|} = e^{-i\beta} \;
\end{equation}
and $\alpha=\pi- \beta -\gamma$,
we have
\begin{eqnarray}
\label{physparam}
e^{i\beta} a_{+-} &=& T^{+-} e^{-i\alpha} + P^{+-}\nonumber\,, \\
e^{i\beta} a_{-+} &=& T^{-+} e^{-i\alpha} + P^{-+}\,, \\
e^{i\beta} a_{00} &=& T^{00} e^{-i\alpha} + P^{00}\nonumber \;.
\end{eqnarray}
The overall weak phase $e^{i\beta}$ is without physical impact
and can be neglected, because in the SM
the weak phase associated with $B^0-\bar B^0$ mixing is controlled
by $q/p=\exp(-2i\beta)$\footnote{Recall that the B mass eigenstates
are defined via $|B_L\rangle = p|B^0 \rangle + q| \bar B^0 \rangle$
and $|B_H\rangle = p|B^0 \rangle - q| \bar B^0 \rangle$. We assume
throughout that the width difference of the two B mass eigenstates
is negligible, so that $|q/p|=1$.}.
Thus $q\bar a_{ij}/p \propto \exp(-i\beta)$, just as $a_{ij}$ is.
Consequently, the isospin analysis in $B\to \rho\pi$ decay
determines $\alpha$.
The crucial assumption of the isospin analysis is to associate
the CKM factor $V_{tb}^\ast V_{td}^{}$ with
$|\Delta I|=1/2$ transitions exclusively, so that from
Eqs.~(\ref{isopm},\ref{isomp},\ref{iso00}), we have
\begin{equation}
\label{isopenguin}
P^{00}=\frac{1}{2}(P^{+-} + P^{-+})\;.
\end{equation}
The overall strong phase in Eq.~(\ref{physparam}) is trivial, so
that with Eq.~(\ref{isopenguin}), we have ten parameters in all,
which can be determined in an analysis of the Dalitz plot~\cite{QS1,QS}.

The presence of the $\sigma\pi$ final state in the phase space
associated with the $\rho^0\pi^0$ channel breaks the relation
assumed in Eq.~(\ref{isopenguin}), and thus mimics the appearance
of isospin violation. In this paper we study how the impact of this
additional decay channel can be minimized. It is worth noting,
however, that the $\sigma\pi$ final state is of definite
CP, so that the isospin analysis can be enlarged to
include this channel as well --- 
additional observables are also present in this case. Before doing this,
let us enumerate the ways in which SM
isospin violation can impact the usual
$B\to \rho\pi$ analysis, to determine whether the impact of
these effects can be reduced as well:
\begin{itemize}
\item[i)] Isospin violation can generate an additional amplitude, of
$|\Delta I|=5/2$ character, as in
Eqs.~(\ref{isopm},\ref{isomp},\ref{iso00}). A $|\Delta I|=5/2$
amplitude can be generated by ${\cal O}(m_d - m_u)$ or
${\cal O}(\alpha)$ effects in concert with a $|\Delta I|=3/2$ weak
transition, or by ${\cal O}(\alpha)$ effects in concert with a
$|\Delta I|=1/2$ weak transition.
The ${\cal O}(m_d - m_u)$ term acts as an isovector interaction.
We recall that the physical neutral pion state is an admixture of
the pseudoscalar octet fields $\pi^0$ and $\eta$; that is,
$(\pi^0)_{\rm phys}= \pi^0 + \epsilon \eta$ with
$\epsilon \sim {\cal O}(m_d-m_u)$. Consequently $\epsilon$
acts as an $I=1$ ``spurion''~\cite{tdlee}, encoding isospin-violating
effects so that the matrix elements with the spurion are $SU(2)_f$
invariant. Isospin violation is also realized via the
$B^+$, $B^0$ mass difference; such effects are not encoded
in the spurion
framework, but they are also comparatively trivial.

\item[ii)] Isospin violation can modify the form factors $f_i$.
The factor $f_0$, e.g., is distinguished by the 
G-parity-violating
decay $\omega \to \pi^+ \pi^-$. The magnitude and phase of this
effective $\rho^0$-$\omega$ ``mixing'' can be elucidated from
$e^+ e^-\to \pi^+\pi^-$ data~\cite{svghbo}; however, the contribution
is reflective of the decay $B^0 \to \omega \pi^0 \to \pi^+\pi^- \pi^0$,
so that $a_{00}$ is modified in this region as well.
In addition,
electromagnetic effects distinguish 
$f_{\mp}$, probed in $\tau$ decay, from 
$f_0$~\cite{ciri,Kubis:1999db}.

\item[iii)] Penguin contributions of $|\Delta I|=3/2$ character can
occur, either through electroweak penguin effects~\cite{randall}, or
through isospin violation in the matrix elements of the
gluonic penguin operator~\cite{svg,Buras:1998ra,svggv1}.

\end{itemize}

The impact of these isospin-violating effects can be
redressed, at least in part. For example, in
$B\to \rho\pi$ decay, the $A_{5/2,2}$
amplitude appears in the combination
$A_{5/2,2}+A_{3/2,2}$ throughout
Eqs.~(\ref{isopm},\ref{isomp},\ref{iso00}). Moreover, the
two amplitudes share the same weak phase, to a good approximation.
This emerges because, unlike $K\to \pi\pi$ decay~\cite{svggv2},
no ``$|\Delta I|=1/2$ rule''
apparently exists in $B\to\pi\pi$ decay ---
though ${\cal B}(B^+\to\pi^+\pi^0)$ has yet to be conclusively
determined~\cite{bpipidata}. Recent theoretical estimates suggest
that the magnitude of the ratio of the $|\Delta I|=1/2$ to $|\Delta I|=3/2$
amplitudes in $B\to\pi\pi$ decay is roughly $0.3$~\cite{BBNS},
so that the $|\Delta I|=5/2$
amplitude is driven by an underlying $|\Delta I|=3/2$
weak transition. 
Strong-interaction isospin violation acts in concert with a
$|\Delta I|=3/2$ weak transition to generate a $|\Delta I|=5/2$
amplitude, whereas electromagnetism can generate a $|\Delta I|=5/2$
amplitude from a $|\Delta I|=1/2$ weak transition.
The size of strong-interaction
isospin violation is typified by the $\pi^0-\eta$ mixing angle
$\epsilon^{(2)}=\sqrt{3}(m_d -m_u)/4(m_s - \hat{m})$ with
$\hat{m}=(m_d + m_u)/2$; we note that
$\epsilon^{(2)} / \alpha \sim 1.45$~\cite{leut96},
enhancing the extent to which
$A_{5/2,2}$ and $A_{3/2,2}$ share the same weak phase.
To the degree that this is true,
the phenomenological $T^{ij}$
parameters of Eq.~(\ref{physparam}) include $|\Delta I|=5/2$
effects as well. 
Thus we see that the single, crucial assumption of the isospin analysis
is that the CKM factor $V_{tb}^\ast V_{td}^{}$ accompanies
$|\Delta I|=1/2$ transitions exclusively, for in this case
the weak phases of the $A_{5/2,2}$ and $A_{3/2,2}$ amplitudes
are identical.
We have shown that
isospin-violating contributions built on the $|\Delta I|=3/2$
short-distance, weak transition do not impact the isospin
analysis in $B\to\rho\pi$. However, non-$|\Delta I|=1/2$
penguin effects,
be they electroweak penguin contributions or contributions
consequent to
isospin-violating effects in the hadronic matrix elements of
$|\Delta I|=1/2$ operators, present a irreducible
hadronic ambiguity from the viewpoint of this analysis.

Empirical information on the all-neutral mode, $a_{00}$, is
essential to the extraction of $\alpha$; however, it is possible
to bound the strong-phase uncertainty using bounds on
$a_{00}$ and its CP-conjugate $\bar a_{00}$~\cite{QS}.
Under the assumptions we have articulated,
the bounds on the hadronic uncertainty
realized in $B\to\rho\pi$ decay~\cite{QS} are not modified by the
presence of a $|\Delta I|=5/2$ transition.

An isospin analysis of $B\to\pi\pi$ decay also permits the
extraction of $\sin(2\alpha)$ from the mixing-induced
CP asymmetry in $B\to \pi^+\pi^-$~\cite{GLo}. In this case, in
contrast,
the $|\Delta I|=5/2$ amplitude cannot be combined with the
$|\Delta I|=3/2$ amplitude. With $A_{|\Delta I|,I_f}$, we have
\begin{equation}
b_{+-}\equiv A(B^0\to\pi^+\pi^-)
= -\frac{1}{\sqrt{3}} A_{1/2,0} + \frac{1}{\sqrt{6}}[A_{3/2,2} - A_{5/2,2}] \;,
\end{equation}
\begin{equation}
b_{00}\equiv A(B^0\to\pi^0\pi^0)
= -\frac{1}{\sqrt{3}} A_{1/2,0} - \sqrt{\frac{2}{3}}[A_{3/2,2} - A_{5/2,2}]\;,
\end{equation}
\begin{equation}
b_{+0}\equiv A(B^+\to\pi^+\pi^0)
=\frac{\sqrt{3}}{2}A_{3/2,2} + \frac{1}{\sqrt{2}}A_{5/2,2} \;.
\end{equation}
As the case of $B\to\rho\pi$, three isospin amplitudes describe
three empirical amplitudes, so that we expect our parametrization
to persist in the presence of isospin violation.
The lowest-dimension operators of the effective weak Hamiltonian
generate transitions of
$|\Delta I|=1/2$ and $|\Delta I|=3/2$ character, so that in the absence of
isospin-violating effects in the hadronic matrix elements,
$A_{1/2,0} \to A_0$ and $A_{3/2,2} \to A_2$, and two amplitudes
suffice to describe the three transitions. The 
isospin analysis in $B\to\pi\pi$ relies on the
relation $(b_{+-} - b_{00})/\sqrt{2} - b_{+0}=0$~\cite{GLo}.
The right-hand side of this relation
is proportional to $A_{5/2,2}$, so that the required relation
is broken by amplitudes of $|\Delta I|=5/2$ character. With
such effects the isospin analysis can fail to determine
the true value of $\sin(2\alpha)$~\cite{svg}.
The small $B\to\pi^0\pi^0$ rate makes the full isospin
analysis difficult to effect, so that
bounds on the hadronic uncertainty in the extraction of
$\sin(2\alpha)$
have also been constructed~\cite{charles,pirjol,GQ,gronau}.
The presence of the $|\Delta I|=5/2$ amplitude,
as well as that of electroweak penguins, imply that the
bounds can underestimate the size of the hadronic uncertainty~\cite{svg}.
However,
bounds which rely on the neutral $B$ modes exclusively,
such as Eq.~(83) of Ref.~\cite{charles},
contain the same linear combination of $|\Delta I|=3/2$ and
$|\Delta I|=5/2$ amplitudes throughout --- so that our
arguments concerning the $|\Delta I|=5/2$ amplitude in $B\to\rho\pi$
decay are germane here as well. 
We conclude, to the extent
the $|\Delta I|=3/2$ and $|\Delta I|=5/2$ amplitudes
share the same weak phase, that
such bounds are insensitive to the $|\Delta I|=5/2$ amplitude
and yield more reliable bounds on the
hadronic uncertainty.

\section{Extension of the isospin analysis: Inclusion of the
\boldmath{$\sigma \pi$} channel}
\label{sec:extend}
The $B\to\sigma\pi$ channel has definite properties under CP,
so that it can be included in the $B\to\rho\pi$ analysis as well.
Defining $a^{\sigma}_{00}=A(B^0\to\sigma\pi^0)$, we have
\begin{equation}
\label{defsig0}
e^{i\beta} a^{\sigma}_{00} = T_{\sigma}^{00} e^{-i\alpha} + P_{\sigma}^{00}\;.
\end{equation}
$T_{\sigma}^{00}$ and $P_{\sigma}^{00}$ are unrelated to the parameters
of Eq.~(\ref{physparam}),
so that we gain four additional hadronic parameters. However, more
observables are present as well.
Including the scalar channel, we now have
$A_{3\pi}\equiv
A(B^0\to \pi^+\pi^-\pi^0)=f_+\, a_{+-} + f_-\, a_{-+} + f_0\, a_{00}
+ f_\sigma \, a^{\sigma}_{00}$, where $f_\sigma$ is the form factor
describing $\sigma\to\pi^+\pi^-$.
It is worth noting that any
discernable presence of the $B\to \sigma\pi$ channel in the
$B\to \rho\pi$ phase space falsifies the notion that
the ``nonresonant'' background can be
characterized by a single, constant
phase across the Dalitz plot~\cite{charles_nr}. For further
discussion of the treatment of nonresonant contributions,
specifically in  $D\to 3\pi$ decay, see Ref.~\cite{frabetti}
--- note also Ref.~\cite{babarbook}.

Neglecting the width difference of the B-meson mass eigenstates, as
$\Delta \Gamma \equiv \Gamma_H - \Gamma_L$ and $|\Delta\Gamma| \ll
\Gamma\equiv (\Gamma_H + \Gamma_L)/2$, we note that the decay
rate into $\pi^+\pi^-\pi^0$ for a $B^0$ meson at time $t=0$ is given
by~\cite{Azimov:ra}
\begin{equation}
\Gamma(B^0(t) \to \pi^+\pi^-\pi^0) = |A_{3\pi}|^2 \exp(-\Gamma t)
\left[\frac{1 + |\lambda_{3\pi}|^2}{2} +
\frac{\left(1 - |\lambda_{3\pi}|^2\right)}{2} \cos(\Delta m\,t)
-{\rm Im}\,\lambda_{3\pi}\sin(\Delta m\,t) \right] \;,
\end{equation}
whereas the decay rate into $\pi^+\pi^-\pi^0$ for a $\bar B^0$
meson at time $t=0$ is given
\begin{equation}
\Gamma(\bar B^0(t) \to \pi^+\pi^-\pi^0) = |A_{3\pi}|^2 \exp(-\Gamma t)
\left[\frac{1 + |\lambda_{3\pi}|^2}{2} -
\frac{\left(1 - |\lambda_{3\pi}|^2\right)}{2} \cos(\Delta m\,t)
+{\rm Im}\,\lambda_{3\pi}\sin(\Delta m\,t) \right]\;.
\end{equation}
We note that
$\lambda_{3\pi} \equiv q \bar A_{3\pi}/p A_{3\pi}$, where we
have defined
$\bar A_{3\pi}\equiv A(\bar B^0\to \pi^+\pi^-\pi^0)$,
and $\Delta m \equiv M_H - M_L$. Different observables are
possible. For example, we can consider
untagged observables, for which the identity of the $B$ meson
at $t=0$ is unimportant, so that
$\Gamma(B^0(t) \to \pi^+\pi^-\pi^0) + \Gamma(\bar B^0(t) \to \pi^+\pi^-\pi^0)
\propto (1 + |\lambda_{3\pi}|^2)$, or we can consider time-integrated,
tagged observables, containing
$\Gamma(B^0(t) \to \pi^+\pi^-\pi^0) - \Gamma(\bar B^0(t) \to \pi^+\pi^-\pi^0)$,
which are sensitive to
$(1 - |\lambda_{3\pi}|^2)$. The products $f_i f_j^*$ contained
therein are distinguishable through the Dalitz plot of this decay
and thus the coefficients of these functions are distinct
observables~\cite{QS1}.
Were we to neglect the $\sigma \pi$ channel,
nine distinct, untagged observables exist, so that all the
hadronic parameters save one would be determinable from the untagged data,
for which greater statistics will be available~\cite{QS}.
If we enlarge the analysis to include the $\sigma\pi^0$ channel,
the additional
interferences possible imply that there are now sixteen distinct,
untagged observables. Moreover, there are fifteen, rather than eight,
tagged, time-integrated observables as well. Nevertheless, it
would seem that the additional hadronic parameters associated
with the $\sigma\pi^0$ final state can be extracted from
untagged data alone. Of course the practicability of the procedure
relies on the amount of data eventually collected; moreover,
the observables are highly correlated.

\section{Evaluating 
\boldmath{$B \to \pi^+\pi^-\pi^0$} decay}
\label{sec:evalB3pi}
The effective, $|\Delta B|=1$ Hamiltonian for
$b\to d q\bar q$ decay is given by
\begin{equation}
{\cal H}_{\rm eff} = \frac{G_F}{\sqrt{2}}
\left[ \lambda_u (C_1 O_1^u + C_2 O_2^u) +
\lambda_c (C_1 O_1^c + C_2 O_2^c) - \lambda_t \sum_{i=3}^{10} C_i O_i
\right] \;,
\end{equation}
where $\lambda_q \equiv V_{qb}^{} V_{qd}^\ast$ with $V_{ij}^{}$
an element of the CKM matrix. The Wilson coefficients $C_i$
and operators $O_i$ are detailed in Ref.~\cite{buras_rmp}, though
we shall interchange $C_1 O_1^{q} \leftrightarrow  C_2 O_2^{q}$ so that
$C_1\sim {\cal O}(1)$ and $C_1 > C_2 \gg C_{3\dots 10}$.
The contributions with $i=3\dots 6$ correspond to strong penguin effects,
whereas those with $i=7\dots 10$ are characterized by
electroweak penguin effects. The Wilson coefficients
with $i=7\dots 10$
are numerically smaller than
those with $i=3\dots 6$, and penguin effects are not
CKM-enhanced in $b\to d q\bar q$ decay, so that we shall
neglect the terms with $i=7\dots 10$ all together.

The decay amplitude for $B\to M_1 M_2$, where $M_1$ and $M_2$ are mesons,
is given by
\begin{equation}
A(B\to M_1 M_2) = \langle M_1 M_2 | {\cal H}_{\rm eff} | B \rangle \;.
\end{equation}
The requisite matrix element contains terms of the form
\begin{equation}
C_i(\mu) \langle M_1 M_2 | O_i | B \rangle \;.
\end{equation}
We adopt the naive factorization approximation
to effect estimates of the hadronic matrix elements. To wit, we
separate $O_i$ into a product of factorized currents, $j_1\otimes j_2$,
and evaluate
$\langle M_1 | j_1 | B \rangle \langle M_2 | j_2 | 0 \rangle$,
so that the operator matrix element becomes a product of a form
factor and a decay constant. Such a treatment, albeit simple, is
incomplete.
The amplitude $A(B\to M_1 M_2)$ is related to a physical observable
and as such must be $\mu$-independent, though the $C_i$ therein
do depend on $\mu$. Evidently the $\mu$ dependence of the
operator matrix elements compensates to yield a
$\mu$ independent result. In the naive factorization approximation,
we have replaced the operator matrix element by a product
of a form factor and decay constant. These quantities are themselves
physical observables and thus are without $\mu$ dependence, so that
the overall $\mu$ dependence of the computed amplitude remains.
Effecting this approximation, however,
allows us to realize a clear connection
to earlier
work~\cite{deA,dearho}, for our purpose is to illustrate the
impact of using a scalar form factor consistent with low-energy
constraints.

In the naive factorization approximation, we can replace
the effective Hamiltonian by the sum of products of
factorized currents, so that
${\cal H}_{\rm eff} = {\cal T}^{1,2} + {\cal T}^{3,4} + {\cal T}^{5,6}$,
where
\begin{equation}
\label{T12def}
{\cal T}^{1,2}=\frac{G_F}{\sqrt{2}} \lambda_u
\left[ a_1 \bar u \gamma^\mu(1-\gamma_5) b \otimes
\bar d \gamma_\mu (1-\gamma_5)u  +
a_2 \bar d \gamma^\mu(1-\gamma_5) b \otimes
\bar u \gamma_\mu (1-\gamma_5)u
\right] \;,
\end{equation}
\begin{equation}
\label{T34def}
{\cal T}^{3,4}=-\frac{G_F}{\sqrt{2}} \lambda_t
\left[ a_3
\sum_q \bar d \gamma^\mu(1-\gamma_5) b \otimes
\bar q \gamma_\mu (1-\gamma_5)q  +
a_4 \sum_q \bar q \gamma^\mu(1-\gamma_5) b \otimes
\bar d \gamma_\mu (1-\gamma_5)q
\right] \;,
\end{equation}
and
\begin{equation}
\label{T56def}
{\cal T}^{5,6}=-\frac{G_F}{\sqrt{2}} \lambda_t
\left[ a_5
\sum_q \bar d \gamma^\mu(1-\gamma_5) b \otimes
\bar q \gamma_\mu (1+\gamma_5)q
-2 a_6 \sum_q \bar q (1-\gamma_5) b \otimes
\bar d (1+\gamma_5)q
\right]\;.
\end{equation}
We define
$a_i\equiv C_i + C_{i+1}/3$ for $i$ odd and
$a_i\equiv C_i + C_{i-1}/3$ for $i$ even; note, too,
that $q\in u,d,s,c$ if $\mu\lesssim m_b$.

We now specifically consider $B\to \sigma\pi$ transitions; the relevant
formulae for $B\to\rho\pi$ are detailed in App.~\ref{app:Brhopi}.
By ``$\sigma$,'' we always mean a two--pion state with total isospin zero and
in a relative S--wave state, $(\pi \pi)_{{\rm S}}$,
understanding its dynamical origin
in the strong pionic FSI
for these quantum numbers --- see the
following section for a more detailed discussion. The matrix elements
involving the $\sigma$ are
\begin{equation}
q^\mu \langle \sigma(p_\sigma) | \bar u \gamma_\mu
(1-\gamma_5) b | \bar B^0(p_B) \rangle=
-i(M_B^2 - M_\pi^2)F_0^{B\to \sigma}(q^2) \;,
\end{equation}
where $q\equiv p_B - p_\sigma$ and
$\langle \sigma | \bar q' \gamma_{\mu} (1+\gamma_5) q'|0\rangle$, noting
$q'\in u,d,s$, {\it vanishes} by C-invariance. It is just such a suppression
mechanism that prompts the authors of Ref.~\cite{hiller} to argue
that non-factorizable effects ought be relatively enhanced in
two-body B decays to final states with scalar mesons,
as the factorization contribution
is itself small. Nevertheless, as our focus is the scalar form
factor itself, we proceed with our estimates.

For the $\pi^+$ we have
\begin{equation}
\label{fpidef}
\langle \pi^+(p) | \bar u \gamma^\mu(1-\gamma_5)d| 0 \rangle= i f_{\pi} p^\mu
\end{equation}
and
\begin{equation}
\label{Btopidef}
\langle \pi^+(p) | \bar u \gamma_\mu (1-\gamma_5) b | \bar B^0(p_B)\rangle
= \left[(p_B+p)_\mu -
\frac{(M_B^2 - M_\pi^2)}{q^2}q_\mu
\right] \frac{F_1^{B\to\pi}(q^2)}{2}
+ \frac{(M_B^2 - M_\pi^2)}{q^2}q_\mu
F_0^{B\to\pi}(q^2)\;.
\end{equation}
With these relations, Eqs.~(\ref{T12def},\ref{T34def},\ref{T56def}),
and the equations of motion for the quark fields,
we have
\begin{eqnarray}
\label{Btosigmapim}
\langle \pi^- \sigma
| {\cal H}_{\rm eff} | B^- \rangle &=&  \frac{G_F}{\sqrt{2}}
\Bigg\{
 f_\pi (M_B^2 - M_\sigma^2) F_0^{B\to\sigma}(M_\pi^2)
\left[ \lambda_u a_1 - \lambda_t a_4 +
\lambda_t \frac{a_6 M_\pi^2}{\hat{m}(m_b+\hat{m})} \right] \\
&+&\lambda_t 2a_6
\frac{\langle \sigma | \bar d d | 0 \rangle}
{(m_b - \hat{m})}
(M_B^2 - M_\pi^2) F_0^{B\to\pi}(M_\sigma^2)
\Bigg\} \nonumber
\end{eqnarray}
and
\begin{eqnarray}
\label{Btosigmapi0}
\langle \pi^0 \sigma
| {\cal H}_{\rm eff} | \bar B^0 \rangle &=&  \frac{G_F}{\sqrt{2}}
\Bigg\{
 \frac{f_\pi}{\sqrt{2}} (M_B^2 - M_\sigma^2) F_0^{B\to\sigma}(M_\pi^2)
\left[ \lambda_u a_2 + \lambda_t a_4 -
\lambda_t \frac{a_6 M_\pi^2}{\hat{m}(m_b+\hat{m})} \right] \\
&-&\lambda_t 2a_6
\frac{\langle \sigma | \bar d d | 0 \rangle}
{(m_b - \hat{m})}
(M_B^2 - M_\pi^2) \frac{F_0^{B\to\pi}(M_\sigma^2)}{\sqrt{2}}
\Bigg\} \;,\nonumber
\end{eqnarray}
where we have replaced the $u,d$ quark masses with $\hat{m}$
and set $M_{\pi^0}=M_{\pi^\pm}=M_{\pi}$
and $M_{B^\pm}=M_{B^0\,,\bar B^0}=M_B$, 
as we neglect isospin-violating
effects. We adopt the usual phase conventions for the flavor wave functions
$\pi^+,\pi^0,\pi^-= u\bar d,(u\bar u - d\bar d)/\sqrt{2},d\bar u$,
and adopt analogous relations for the $\rho$ mesons as well.
In the context of the $B\to\rho\pi$ analysis, the decay
$\bar B^0 \to \sigma\pi$, specifically its penguin contributions,
modify Eq.~(\ref{isopenguin}), the assumption crucial to
the analysis. Thus it is important to make an 
assessment of the size of penguin effects in this decay.
The terms containing $a_6$, the scalar penguin contribution,
are formally $1/m_b$ suppressed, but can be chirally enhanced:
the numerical factor
$M_{\pi}^2/(\hat{m}m_b)\sim 0.6$ is only modestly less than
unity. The second term proportional to $a_6$ contains
$\langle \sigma | \bar d d | 0 \rangle$, where
we anticipate, in the vicinity of the $\sigma$ resonance,
$\Gamma_{\sigma\pi\pi}(s)\langle \sigma | \bar d d | 0\rangle
=\langle \pi^+(p_+)\pi^-(p_-) | \bar d d |0 \rangle = \sqrt{2/3}B_0
\Gamma_1^{n\,\ast}(s)$,
where $\Gamma_1^n(s)$ is the scalar form factor, which we detail in the
next section, and $s=(p_+ + p_-)^2$. The parameter
$B_0$ is related to the vacuum quark condensate.
Neglecting small terms of second order in the quark masses,
$B_0 = - \langle 0|\bar q q|0\rangle / F_{\pi}^2$,
where $F_{\pi}$, the $\pi^0$ decay constant, is $f_\pi/\sqrt{2}$.
Commensurately, we can simply set $B_0\equiv M_\pi^2/(2\hat{m})$
to realize our numerical estimates.
Note that
$\Gamma_{\sigma\pi\pi}$ describes the $\sigma\to\pi^+\pi^-$ form factor.
With our conventions,
$B_0>0$, so that the two $a_6$ contributions are of the same sign.
Using the parameters of Ref.~\cite{deA}
and $\Gamma_{\sigma\pi\pi}=\Gamma_1^{n\,\ast}\chi$ with,
as we shall determine, $\chi=20.0\,{\rm GeV}^{-1}$,
we find that the $a_6$ term containing $f_\pi$ 
to be roughly a factor of four larger.
The
$\langle \sigma | \bar d d| 0 \rangle$ term, present in the penguin
contributions in $B\to\sigma\pi$, slightly enhances the $a_6$
contribution and its subsequent
cancellation with the $a_4$ contribution,
as $a_4$ and $a_6$ are of the same sign. The same cancellation occurs
in  $\bar B^0\to \rho^0\pi^0$ (see
App. A for a compilation of the relevant formulae), so that
penguin contributions in $\bar B^0\to \sigma\pi^0$ can be expected
to be crudely comparable
to those in $\bar B^0 \to \rho^0\pi^0$.
Modifications of Eq.~(\ref{isopenguin}) can thus be expected to occur.

In the previous section we determined how the isospin analysis
in $B\to\rho\pi$ could be extended to include $B\to \sigma\pi$.
Eqs.~(\ref{Btosigmapim}) and (\ref{Btosigmapi0}), however, are
related by isospin symmetry, so that 
it is useful to determine whether
additional constraints
on $P_{\sigma}^{00}$ in Eq.~(\ref{defsig0}) can be realized.
Parametrizing the amplitudes in terms
of $A_{|\Delta I|,I_f}$, we have
\begin{equation}
\label{isosig00}
A(B^0\to \sigma\pi^0) =
\frac{1}{\sqrt{2}} A_{1/2,1}  - \frac{1}{\sqrt{2}} A_{3/2,1}
\end{equation}
and
\begin{equation}
\label{isosig0p}
A(B^+\to \sigma\pi^+) =
- A_{1/2,1}  - \frac{1}{2} A_{3/2,1} \;,
\end{equation}
so that two isospin amplitudes appear for the two empirical amplitudes.
Although the $|\Delta I|=1/2$ amplitudes are simply related, we see
that no useful constraint emerges, as the tree amplitude in
$B^+\to \sigma\pi^+$, which includes $|\Delta I|=1/2$ and $3/2$ amplitudes,
gives rise to two additional, undetermined hadronic parameters.

We have illustrated that the penguin relation essential to the
isospin analysis in $B\to\rho\pi$, Eq.~(\ref{isopenguin}), can be broken
through the presence of the $B\to \sigma\pi$ decay channel. In
our numerical estimates, however, we neglect penguin contributions,
in order to retain a crisp comparison with earlier
work~\cite{deA}, for our purpose is to illuminate the impact of
the $\sigma\to\pi^+\pi^-$ form factor.
With such an approximation, a computation of the Wilson coefficients
in leading order in $\alpha_s$ suffices~\cite{buch_eps}, so that
$C_1(\mu)=1.124$ and $C_2(\mu)=-0.271$ at $\mu=m_b=4.8$ GeV
as per Ref.~\cite{keum}, to yield $a_1=1.034$ and $a_2=0.104$. In contrast,
Ref.~\cite{deA} uses the 
``fitted'' values
$C_1(m_b)=1.105$ and $C_2(m_b)=-0.228$, to yield $a_1=1.029$ and
$a_2=0.140$. These are very similar to the Wilson coefficients
in next-to-leading-order QCD, after Ref.~\cite{buraswc},
used in the $B\to\rho\pi$ analysis of
Ref.~\cite{dearho}.
For definiteness, we shall adopt these last values, detailed in
Sec.~\ref{sec:res}, in
our numerical analysis. The values of $a_1$ are quite similar, whereas
those of $a_2$ differ by tens of percent. Generally, we expect our
numerical predictions for decay channels controlled by $a_2$ to
be less robust, as
the scale dependence of $a_2$,
as illustrated in Table III of Ref.~\cite{keum}, is severe.
Note that it persists to a significant degree
in the next-to-leading order treatment of
Ref.~\cite{BBNS} as well.

Nevertheless, let us proceed to consider numerical estimates for
$B\to\sigma\pi$ decay. We reconstruct the $\sigma$ meson from the
$(\pi^+\pi^-)_{\rm S}$ final state, so that we have
\begin{equation}
\label{Asigpipi}
A_\sigma (B\to \pi^+\pi^-\pi) \equiv
\langle (\sigma\to \pi^+\pi^-)\pi | {\cal H}_{\rm eff} | B \rangle =
A(B\to \sigma \pi) \Gamma_{\sigma\pi\pi}\;,
\end{equation}
where $\Gamma_{\sigma\pi\pi}$ is the $\sigma \to \pi^+\pi^-$
form factor introduced previously.
The scalar form factor contains the meson loop function and thus
a regularization scale $\mu_r$ at which it is evaluated. In our approach,
this scale dependence is disjoint from that
associated with the renormalization
of the operators of the effective, weak Hamiltonian, so that it
is chosen for convenience and is quite independent of $\mu$.
It may seem untoward to graft two very different calculations, namely, of
$A(B\to \sigma \pi)$ and of $\Gamma_{\sigma\pi\pi}$, to yield
$A(B\to (\sigma\to \pi^+\pi^-)\pi)$.
In a holistic treatment one might hope to
recast a resonance and its subsequent decay products in terms
of a single, complex hadron distribution amplitude, $\phi(x,\mu)$, which
describes the non-perturbative dynamics. The
analysis of the decay amplitude could then proceed via standard
pQCD techniques~\cite{BF,BL}. In this manner the $\mu_r$ dependence of which
we have spoken is connected, albeit loosely, to
the scale dependence of $\phi(x,\mu)$. Nevertheless, the
explicit ``QCD factorization'' analysis of Ref.~\cite{BBNS} shows that
the scale dependence of the hadron distribution functions is trivial
in an NLO analysis in $\alpha_s$,  so that the consistency issues
to which we have alluded are beyond the scope of current calculations.

\begin{figure}[htb]
\begin{center}
\epsfig{file=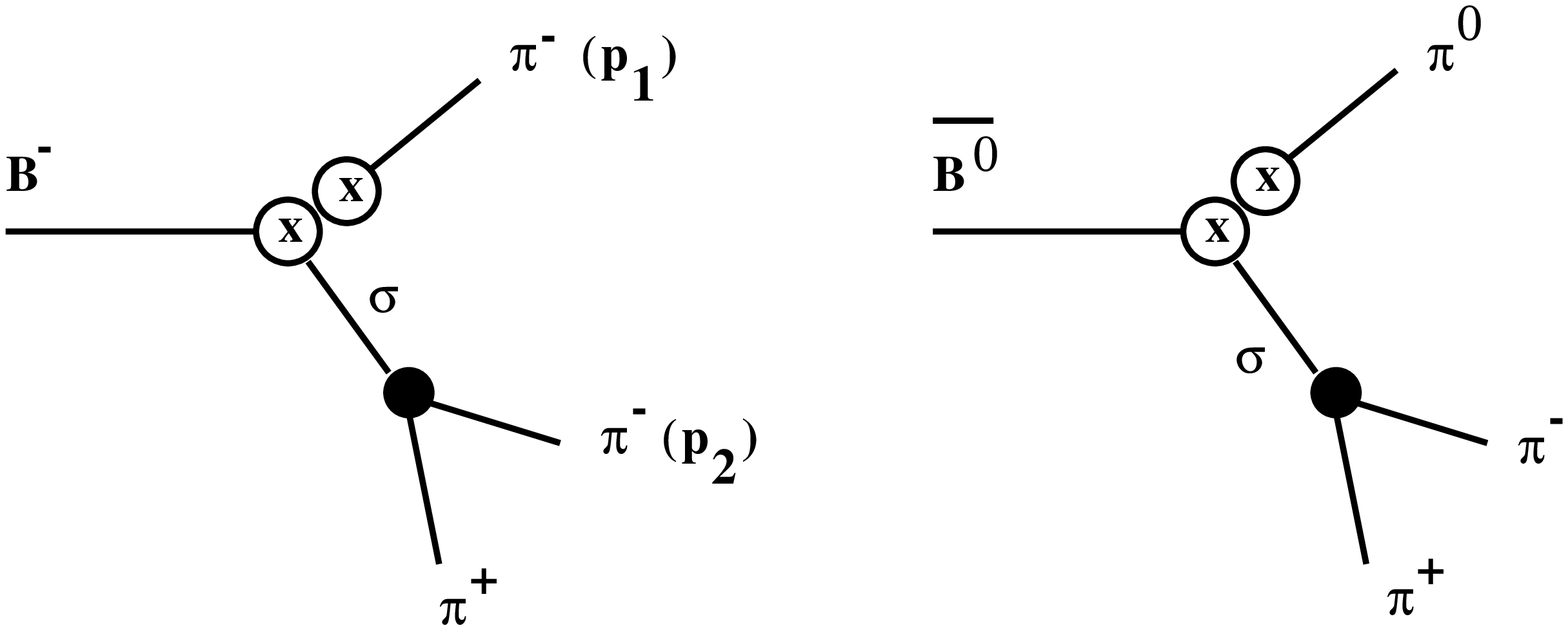, height = 5cm}
\end{center}
\centerline{\parbox{12cm}{\caption{
$B\to\pi^+\pi^-\pi$ decay as mediated by the $\sigma$ resonance.
The factorized weak vertex is denoted by ``$\otimes\otimes$''.
The filled circle denotes the strong three-meson
vertex, here $\sigma \to 2\pi$.
\label{fig:fb1}
}}}
\end{figure}

Turning to $B^- (p_B) \to \pi^+ (p_+)
\pi^- (p_1) \pi^- (p_2)$ decay, we 
define $u= (p_B - p_1)^2= (p_+ + p_2)^2$ and $t =(p_+ + p_1)^2$.
The contributions driven by the $\sigma$ resonance are of the form
$B^- (p_B) \to (\sigma \to \pi^+ (p_+)
\pi^- (p_1)) \pi^- (p_2)$ or
$B^- (p_B) \to (\sigma \to \pi^+ (p_+)
\pi^- (p_2)) \pi^- (p_1)$ --- the latter is illustrated in Fig.~\ref{fig:fb1}.
(The corresponding contributions to $B\to\rho\pi$ decay are illustrated in
Fig.~\ref{fig:fb2}.)
The two contributions add coherently, so that the branching ratio
for $B^- (p_B) \to \pi^+ (p_+)
\pi^- (p_1) \pi^- (p_2)$ is enhanced through the presence of two
identical pions in the final state. In $\bar B^0$ decay this does
not occur, and we have
$\bar B^0 (p_B) \to \pi^+ (p_+)\pi^- (p_1) \pi^0 (p_2)$.
Thus we find
\begin{eqnarray}
\label{cfdeasig}
\langle (\sigma \to \pi^+ \pi^-)\, \pi^- \, | \, {\cal H}_{\rm eff} \,| \,
B^- \rangle &=& \frac{G_F}{\sqrt{2}} V_{ub}^{\ast} V_{ud}^{} \, a_1 \,
F_0^{(B\to \sigma)} (M_\pi^2) (M_B^2 - M_\sigma^2) f_\pi \,
\left[ \Gamma_{\sigma  \pi\pi} (t) + \Gamma_{\sigma  \pi\pi} (u)
\right]~,\\
\label{cfdeasig0}
\langle (\sigma \to \pi^+\pi^-)\, \pi^0 \, | \, {\cal H}_{\rm eff} \,| \,
\bar{B}^0 \rangle &=& \frac{G_F}{\sqrt{2}} V_{ub}^{\ast} V_{ud}^{} \, a_2 \,
F_0^{(B\to \sigma)} (M_\pi^2) (M_B^2 - M_\sigma^2) \frac{f_\pi}{\sqrt{2}} \,
\Gamma_{\sigma  \pi\pi} (t)~.
\end{eqnarray}
In Ref.~\cite{deA},
the $\sigma \to \pi^+ \pi^-$ vertex function is chosen to be
\begin{equation}
\label{GammadeA}
\Gamma_{\sigma  \pi\pi} (x) =
{g_{\sigma \pi^+  \pi^-} } \left(
\frac{1}{x-M_\sigma^2 + i\Gamma_\sigma (x) M_\sigma}
\right)\;,
\end{equation}
where the running width $\Gamma_\sigma(x)$ is defined as
\begin{equation}
\Gamma_\sigma (x) = \Gamma_\sigma \frac{M_\sigma}{\sqrt{x}}
\frac{\sqrt{x/4-M_\pi^2}}{\sqrt{M_\sigma^2/4 -M_\pi^2}} ~.
\end{equation}
We shall adopt, however, the definition
\begin{equation}\label{Gammaus}
\Gamma_{\sigma  \pi\pi} (x) =
{\chi } \, \Gamma^{n\,\ast}_1 (x)~,
\end{equation}
where the normalization $\chi$ is fixed to be identical
to that of Eq.~(\ref{GammadeA}), namely
\begin{equation}
\label{gammanorm}
\chi \, \left| \Gamma_1^n (M_\sigma^2) \right| =
\frac{g_{\sigma \pi^+ \pi^-}}{\Gamma_\sigma (M_\sigma^2) M_\sigma}~.
\end{equation}
Using the parameters $g_{\sigma\pi^+\pi^-}$, $M_\sigma$, and
$\Gamma_\sigma$ of Ref.~\cite{deA}, that is,
$M_\sigma=478\pm 24$ MeV
and $\Gamma_\sigma=324\pm 41$ MeV as reported
by the E791 collaboration in $D^+\to 3\pi$, and $g_{\sigma\pi\pi}=2.52$ GeV,
we find
$\chi=20.0\,\hbox{GeV}^{-1}$. Alternatively,
$\chi=\sqrt{2/3}B_0/\langle \sigma | \bar d d | 0 \rangle$,
so that this procedure determines $\langle \sigma | \bar d d | 0 \rangle$
as well.

The $\sigma$ meson is a broad $I=J=0$ enhancement, close to the
$\rho$ meson in mass, so that $B\to\sigma\pi$ decay 
can contribute to the allowed
phase space of $B\to\rho\pi$ decay as well.
To ascertain the impact of the $B\to\sigma\pi$ channel to $B\to\rho\pi$ decay,
we combine the decay channels at the amplitude level
and then integrate over the relevant three-body phase space
to determine the {\it effective} $B\to\rho\pi$ branching ratio.
For $B(p_B)\to \pi^+(p_+)\pi^-(p_1)\pi(p_2)$ decay, we
define
\begin{equation}
\label{deftheta}
\cos \theta = \frac{{\mathbf p_1^\prime}\cdot {\mathbf p_2^\prime}}
{|{\mathbf p_1^\prime}||{\mathbf p_2^\prime}|} \;,
\end{equation}
where the primed variables refer to the momenta in the rest frame
of the $\pi^+ (p_+^\prime) \pi^- (p_1^\prime)$ pair, so that
$\mathbf{p_1^\prime} + \mathbf{p_+^\prime}=0$. Letting
$p_2^2=M_2^2$, we have
\begin{equation}
\label{phasespace}
\Gamma \left( B\to
\rho_{\rm eff}
\pi \right) = \int_{-1}^{1} d \cos
\theta \, \int_{(M_\rho-\delta)^2}^{(M_\rho+\delta)^2} dt \,
\frac{1}{32(4\pi M_B)^3} \, \left( M_B^2 -t - M_2^2\right)
\, \beta_1' \beta_2' \, \left| {\cal M} \right|^2\;,
\end{equation}
where ``$\rho_{\rm eff}$'' is determined by
$(M_\rho-\delta)^2 \le t \le (M_\rho+\delta)^2$,
$\beta_i^\prime$ refers to the velocity of particle $i$ in
the primed frame, and ${\cal M}$ is
the sum of the amplitudes of interest\footnote{For completeness,
we note that
$u=M_+^2 + M_2^2
+ 2E_+^\prime E_2^\prime(1 + \beta_1^\prime \beta_2^\prime \cos\theta)$,
with $E_+^\prime=(t + M_+^2 - M_1^2)/(2\sqrt{t})$,
$E_1^\prime=(t + M_1^2 - M_+^2)/(2\sqrt{t})$, and
$E_2^\prime=(M_B^2 - t - M_2^2)/(2\sqrt{t})$, where
$p_+^2=M_+^2$ and $p_1^2=M_1^2$.}. Typically $\delta\sim (1-2)\Gamma_\rho$.

The scalar form factor we adopt describes the $f_0(980)\to\pi^+\pi^-$
vertex function as well, albeit with a new normalization factor, as
$\Gamma_{f_0\pi\pi}(t)\langle f_0 | \bar d d | 0\rangle
=\langle \pi^+(p_+)\pi^-(p_1) | \bar d d |0 \rangle = \sqrt{2/3}B_0
\Gamma_1^{n\,\ast}(t)$, in the vicinity of the $f_0(980)$ resonance.
The appearance of the $f_0(980)$ resonance in the $\pi^+\pi^-$ channel
is complicated by the opening of the $K\bar K$ threshold; the
$f_0(980)\to\pi^+\pi^-$ form factor is decidedly not of Breit-Wigner
form. It should be noted that the unitarization procedure we employ
here neglects the $\eta\eta$ channel, though, as we discuss in
the next section, this has
no impact on the description of $\pi\pi$ scattering
below $s\simeq 1.1$ GeV. Note, too, that although multiparticle
final states, particularly the $4\pi$ state, can contribute, they are
demonstrably small for $s\leq 1.4$ GeV~\cite{Bugg:jq,Abele:fr}.
The form factor(s) we adopt can be tested though
the {\it shape} of the $f_0(980)$ contribution in
$B\to f_0(980)\pi \to 3\pi$, as well as  through that in
$B\to f_0(980)\pi \to K^+K^-\pi$.

\section{Coupled-channel pion and kaon scalar form factors}
\label{sec:sff}

In the preceding section, we encountered  the non-strange,
scalar form factor of the pion, $\Gamma_1^n (t)$. Such scalar form
factors play a unique role in strong-interaction physics
because they measure the strength of the quark mass term
${\cal H}_m^{\rm QCD} = m_u \bar uu + m_d \bar dd+ \ldots$, i.e., the
explicit chiral-symmetry-breaking term in QCD. However, since no
scalar-isoscalar sources exist, these form factors
cannot be determined directly but, rather, must be
inferred indirectly from hadron-hadron scattering data.
The most prominent
example in this context is the pion-nucleon sigma term, which can be
extracted from the analytically-continued and Born-term-subtracted
isoscalar S-wave amplitude. Similarly, the pion scalar form factor
can be obtained from meson-meson scattering data, with the added
complication of channel couplings above the $\bar KK$ threshold at
$\sqrt{s} \simeq 1$~GeV.
At lower energies, the pion scalar form factor can be calculated
in CHPT at one- \cite{GL84} and two-loop
accuracy \cite{GMII,BCT}.
The one-loop representation fails at surprisingly low energies,
a consequence of the strong pionic FSI present
in this channel~\cite{truong}. This is also signalled by
the very large pion scalar radius, $\langle r_S^2\rangle_\pi \simeq
0.6\,$fm$^2$, which is sizeably bigger than the pion vector radius
governed by the rho mass $\langle r_V^2\rangle_\pi \simeq 6/M_\rho^2
\simeq 0.4\,$fm$^2$, indicating a smaller breakdown scale in the
scalar-isoscalar channel. In fact, these FSI are so strong that they can
dynamically generate the 
$\sigma$ meson, with no need of a genuine
(pre-existing) quark model state~\cite{OO}.
Furthermore, extending such an analysis
to three flavors, one finds that all the scalar mesons with mass
below 1.1~GeV can be dynamically
generated, so that the genuine quark model nonet would have its center of
gravity at about 1.3~GeV, see e.g. Ref.~\cite{JOP} and references therein.
Irrespective of
this admittedly controversial assignment of the
scalars, for further recent discussion see Ref.~\cite{cherry},
the pion scalar form factor
{\bf cannot} be represented simply in terms of a scalar meson with a certain
mass and width.
Although the {\it description} of the low-lying scalar states
in terms of dynamically generated, rather than ``pre-existing,'' states may
be controversial and thus subject to ongoing discussion, let us stress
that the scalar form factor itself is not. The form factor which
emerges for the chiral unitary approach adopted here is quite comparable
to that which emerges from the dispersion analysis of Ref.~\cite{DGL}.
Consequently, when one has a 
source with vacuum quantum numbers coupled to a two-pion state, it
can be very misleading to use a simple Breit-Wigner parametrization,
albeit with a running width. We shall display this graphically
after discussing the scalar form factor and its
construction, to be used in calculating the branching ratios to which
we have alluded.
We note in passing that this casts doubt on the extraction
of the $\sigma$ meson properties from $D\to 3\pi$ decays, note
Ref.~\cite{e791ex}. This is quite in contrast to the pion vector form factor
and the $\rho$ meson, for which such a description works to 
good accuracy.

We now summarize how to calculate the scalar form factor in
Eq.~(\ref{Gammaus}), following Ref.\cite{OM}, to which we
refer in all details.
Of course, one could also use the numerical result of the dispersion
analysis of \cite{DGL} for the scalar form factor. Note that the
band found there overlaps with the band of Ref.~\cite{OM} if one allows
for parameter variations within known bounds. As a first step, we
prefer to work in the framework of the chiral unitary approach, as
it yields a form factor which is more convenient for applications.
As we have noted, we need the scalar form
factor of the pion for momentum transfers larger than
1~GeV. Consequently, we cannot work with the scalar form factor
as computed in CHPT, but must invoke some
resummation technique, as well as account for the channel coupling
between the $\pi \pi$ and the $K\bar K$ systems. The resummation method
is constrained only by unitarity
and thus is not entirely model-independent. However,
it can be strongly constrained by requiring
that the so-constructed
form factors match to the CHPT expressions,
in the region where CHPT is applicable. Due to the
channel coupling, we have to consider transition matrix elements of the
non--strange and strange  scalar quark bilinears between two meson states
of isospin zero, namely $\pi\pi$ and $\bar K K$, and the
vacuum. We exclude the $\eta\eta$ channel, as it
does not affect the phase shifts or pion/kaon decays and
transitions --- it only plays a role in describing
the inelasticity of $I=J=0$ $\pi\pi$ scattering above 1.1~GeV, the
physical $\eta\eta$ threshold.
For a detailed
discussion of this point, we refer to Refs.~\cite{OO,GO,NK}.
The pertinent matrix elements  are given in
terms of  four {\sl scalar} form factors\footnote{Here
``$\pi\pi$'' and ``$KK$'' states denote
the linear combinations of physical $\pi\pi$ and $KK$ states, respectively,
with zero total isospin.},
\ba
\label{rest2}
\langle0|\bar{n}n|\pi\pi\rangle&=&\sqrt{2}\,B_0\, \Gamma^n_1(s)~, \quad
\langle0|\bar{n}n|K\overline{K}\rangle = \sqrt{2}\, B_0 \, \Gamma^n_2(s)~,
\nonumber \\
\langle0|\bar{s}s|\pi\pi\rangle&=&\sqrt{2} B_0 \, \Gamma^s_1(s) ~,\quad
\langle0|\bar{s}s|K\overline{K}\rangle = \sqrt{2}\,B_0 \,
\Gamma^s_2(s)~,
\ea
where $B_0$ is a measure of the vacuum quark condensate, $B_0 =
-\langle 0| \bar q q |0\rangle/F_\pi^2$, with $F_\pi \simeq 93\,$MeV the
pion decay constant (strictly speaking its value in the 
limit of vanishing quark masses). In Eq.(\ref{rest2}),
the following notation is employed. The superscript $s(n)$ refers to the
strange/non--strange quark operator whereas the subscript
$1(2)$ denotes pions
and kaons, respectively. Furthermore,
$\bar{n}n = (\bar{u}u+\bar{d}d)/\sqrt{2}$.
The pion scalar form factors $\Gamma^n_1(s)$ and $\Gamma^s_1(s)$ are
calculated in Ref.\cite{GL84,GL85} through
one loop in CHPT and $\Gamma^n_1(s)$
through two loops in Refs.~\cite{GMII,BCT}.
The scalar kaon form factors at next-to-leading order
in CHPT were first explicitly given in \cite{OM}.
Since it is central to our discussion, we give the explicit
expression for $\Gamma^n_1(s)$~\cite{GL85}
\ba
\label{glpi}
\Gamma^n_1(s)&=&\sqrt{\frac{3}{2}}\left\{1+\mu_\pi-\frac{1}{3}\mu_\eta
+\frac{16 M_\pi^2}{F_\pi^2}
\left(2L_8^r-L_5^r\right)+8(2 L_6^r -L_4^r)\frac{2
M_K^2+3M_\pi^2}{F_\pi^2}+f(s)+\frac{2}{3}\,\widetilde{f}(s)\right\}~,
\ea
with $f(s)$ and  $\widetilde{f}(s)$  given by
\ba
\label{ft}
f(s)&=&\frac{2s-M_\pi^2}{2F_\pi^2}\bar{J}_{\pi\pi}(s)-\frac{s}{4F_\pi^2}\bar{J}_{KK}(s)-
\frac{M_\pi^2}{6F_\pi^2}\bar{J}_{\eta\eta}(s)+
\frac{4s}{F_\pi^2}\left\{L_5^r-\frac{1}{256\pi^2}\left(4\log
\frac{M_\pi^2}{\mu^2}-\log\frac{M_K^2}{\mu^2}+3 \right) \right\}~,\nonumber \\
\widetilde{f}(s)&=&\frac{3}{4}\,\frac{s}{F_\pi^2}\bar{J}_{KK}(s)+\frac{M_\pi^2}{3F_\pi^2}
\bar{J}_{\eta\eta}(s)+\frac{12s}{F_\pi^2}\left\{L_4^r-\frac{1}{256
\pi^2}\left(\log
\frac{M_K^2}{\mu^2}+1 \right)\right\}~.
\ea
Here, $\bar{J}_{PP}(s)$ ($P=\pi,K,\eta$) is
the standard meson loop function~\cite{GL84},
and $\mu$, in this section, is the scale of dimensional regularization.
The quantities $\mu_P$ in Eq.(\ref{glpi}) are given by
\be
\mu_P=\frac{m_P^2}{32\pi^2 F_\pi^2}\log\frac{m_P^2}{\mu^2}~.
\ee
Furthermore, the $L_i^r (\mu )$ are scale--dependent, renormalized
low--energy constants. We use here the same values as in Ref.~\cite{OM}.
Since the combination $2L_6^r - L_4^r$ multiplies
$M_K^2$, the normalization of $\Gamma_1^n(s)$ is
sensitive to the precise value of these low-energy constants, which
are only poorly known. This motivates our choice for the normalization of
$\Gamma_{\sigma\pi\pi}(s)$, given in Eq.~(\ref{gammanorm}).
The one--loop representation can not be trusted beyond
$\sqrt{s} \simeq 400\,$MeV, as 
it begins to diverge from the form factor extracted from $\pi\pi$
scattering data.
To go to higher energies, one therefore
has to study the constraints that unitarity imposes on the scalar form
factors.  The imaginary part of any scalar form factor is
given by the appropriate  meson--meson scattering
$T$--matrix, so that the starting point for any unitary
resummation scheme is exactly this scattering $T$-matrix~\cite{npa},
\be
\label{t}
T(s)=\left[I+K(s)\cdot g(s) \right]^{-1}\cdot K(s)  ~,
\ee
where $s$ denotes the center-of-mass
energy squared and $K(s)$ can be obtained
from the lowest order CHPT Lagrangian, e.g.,
$K(s)_{11}= (s-M_\pi^2/2)/F_\pi^2$. This $T$-matrix not only describes
meson-meson scattering data but also, after gauging, photon decays and
transitions, as reviewed in Ref.~\cite{ollerrev}.
In Eq.(\ref{t}), the diagonal matrix $g(s)$
is nothing but the familiar scalar loop integral
\be\label{gis}
g(s)_i = \frac{1}{(4\pi)^2} \biggl( -1 + \log{\frac{M_i^2}{\mu^2}} +
\sigma_i (s) \log{\frac{\sigma_i (s) + 1}{\sigma_i (s) -1}} \biggr)~,
\ee
given here in dimensional regularization for the $\overline{\rm MS}$
scheme, and $\sigma_i (s) = \sqrt{1-4M_i^2/s}$.
In what follows, we will set the regularization scale $\mu = 1.08\,$GeV.
Of course, observables do not depend on
this choice, and we could choose another value for $\mu$. However, the
original investigation of meson-meson scattering transitions and
decays by Oller and Oset~\cite{npa} uses a
three-momentum cut-off in the pion loop function. Translated to
dimensional regularization, this gives the stated value of $\mu$.
For energies above the threshold of
the state $i$, unitarity implies the following
relation between the form factors and the isospin-zero scattering $T$-matrix,
\be
\label{uni2}
\hbox{Im}~\Gamma(s)=T(s)\cdot \frac{Q(s)}{8\pi\sqrt{s}}\cdot \Gamma^*(s) \;,
\ee
employing an obvious matrix notation with
\begin{equation}
Q(s)=\left(\matrix{p_1(s)
\theta(s-4M_1^2) & 0 \cr 0 & p_2(s)\theta(s-4M_2^2)}\right)~,
\quad \Gamma(s)=\left(\matrix{\Gamma_1(s) \cr \Gamma_2(s)}\right)~,
\end{equation}
% update to use ams package
%\begin{equation}
%Q(s)=\begin{pmatrix} p_1(s)\theta(s-4M_1^2) & 0 \\
%0 & p_2(s)\theta(s-4M_2^2) \end{pmatrix} ~,
%\quad \Gamma(s)=
%\begin{pmatrix} \Gamma_1(s) \\ \Gamma_2(s)\end{pmatrix}~,
%\end{equation}
where $p_i(s)=\sqrt{s/4-M_i^2}$ is the modulus of the c.m.
three-momentum of the state $i$.  Substituting $\hbox{Im}~\Gamma(s)$
with
$(\Gamma(s)-\Gamma(s)^*)/(2i)$ and $T(s)$ with the expression of
Eq.~(\ref{t}) and using the properties of the matrices $g(s)$ and
$K(s)$, one can express $\Gamma(s)$ as:
\be
\label{gamma}
\Gamma(s)=\left[I+K(s)\cdot g(s)\right]^{-1} \cdot R(s)  \;,
\ee
where $R(s)$ is
a vector of functions free of any singularity. We remark that
this procedure of taking the final state interactions
into account is based on
the work of Ref.~\cite{basde}. We also wish to stress that
Eq.(\ref{gamma}) can be applied to any K-matrix
without unphysical cut contributions.
As the final step, one  fixes  the functions in the vector
$R(s)$ by requiring  matching of Eq.(\ref{gamma}) to the
next-to-leading order (one loop)
CHPT $\pi\pi$ and $K\bar{K}$ scalar form factors.
This  matching
ensures that for
energies where CHPT is applicable, these form factors fulfill
all requirements
given by chiral symmetry and the underlying power counting. The
matching procedure thus determines the vector $R(s)$,
as detailed in Ref.\cite{OM}. For completeness, we give
the expression for $R^n_1$ pertinent to the scalar form
factor $\Gamma_1^n (s)$,
\ba
\label{R}
R^n(s)_1 = \sqrt{\frac{3}{2}}\Bigg\{1+\frac{4(L_5^r+2
L_4^r)}{F_\pi^2}s+\frac{16
(2L_8^r-L_5)}{F_\pi^2}M_\pi^2+\frac{8(2L_6^r-L_4^r)}{F_\pi^2}
(2M_K^2+3M_\pi^2)
 - \frac{M_\pi^2}{32\,\pi^2\,F_\pi^2}-\frac{1}{3}\mu_\eta\Bigg\} ~,
\ea
where we have used the Gell-Mann--Okubo relation $3M^2_\eta=4
M_K^2 - M_\pi^2$. This representation of the scalar form factors
is valid from threshold up to energies of about 1.2~GeV. This range
could be extended to higher energies by including multi--particle states. In
fact, as we have noted before, in the u--channel one encounters larger
values of $\sqrt{s}$. Therefore, we simply match
our representation at $\sqrt{s_0} = 1.2 \,$GeV
to the following asymptotic forms
\be
{\rm Re}~\Gamma_1^n (s) \to \frac{a}{s}~, \quad
{\rm Im}~\Gamma_1^n (s) \to \frac{b}{s^2}~, \quad
s \to \infty~.
\ee
We have checked that the final results are insensitive to this
choice of the matching point. The asymptotic form of the real
part of the scalar form factor follows from quark counting
rules~\cite{qcr} in the crossed channel; it
has also  been found in the
dispersion analysis of the Higgs
decay into two pions \cite{DGL}.
For continued contact with Ref.~\cite{deA} we match to the
asymptotic form of the vertex function of Eq.~(\ref{GammadeA});
a more precise treatment, if it were warranted, would involve
smoothly letting $K(s)\to 0$ and solving for the form factors
in a manner consistent with unitarity --- for further discussion,
see Ref.~\cite{DGL}.
In Fig.~\ref{fig:sff} we display the scalar form factor $\Gamma_1^n
(s)$ for $\sqrt{s} \leq 1.2\,$GeV.
In the low--energy region, the modulus of the form factor has
it maximum at $\sqrt{s} \simeq 0.46\,$GeV, very close to the
central value of the $\sigma$ meson mass deduced by the E791
collaboration from analyzing the $D\to 3\pi$ data.
In fact, since we wish to examine the consequences
of using the more general vertex, Eq.(\ref{Gammaus}), as compared
to the choice Eq.(\ref{GammadeA}), used by Ref.~\cite{deA},
we have fixed the normalization
constant $\chi$ such that ${\rm Re}~\Gamma_1^n (s)$ has the
same value as the Breit--Wigner representation 
at $\sqrt{s} = 0.478\,$GeV. This amounts to setting $\chi =
20.0\,$GeV$^{-1}$. The peak at $\sqrt{s} \simeq
1\,$GeV is due to the $f_0 (980)$ and the opening of the $\bar K K$
channel.
Also shown in Fig.~\ref{fig:sff}
is the scalar form factor generated using the Breit--Wigner form
with a running width, adopted in Refs.~\cite{e791ex,deA}.
The differences between this form factor and that
deduced from the low--energy effective
field theory of QCD are marked. In particular, in
Fig.~\ref{fig:sff} we see that the Breit-Wigner representation
of $\Gamma_{\sigma\pi\pi}$
is deficient in that i) $\hbox{Im}\,\Gamma_{\sigma\pi\pi}(s)$
has a different shape as $s$ approaches physical threshold,
$s\to 4M_\pi^2$ and
ii) $\hbox{Re}\,\Gamma_{\sigma\pi\pi}(s)$ does not possess a
unitarity cusp at $s=4M_\pi^2$. It is also of the wrong sign in this limit.
Moreover, the shapes of the
two form factors are very different above $\sqrt{s}\simeq 0.5\,$GeV --- this
has particular consequence for the $B\to\rho\pi$ analysis.

\medskip

\section{Vector form factor}
\label{sec:vector}

Thus far we have considered the $\sigma$ meson contribution to
$B \to \pi^+ \pi^- \pi$ decay. In this section we turn to
the $\rho$ meson contribution,
$B(p_B)\to \rho \pi \to \pi^+(p_+)\pi^-(p_1)\pi(p_2)$,
which presumably dominates for
$t\simeq M_\rho^2$.
In analogy to Eq.~(\ref{Asigpipi}),
the amplitude for $B\to \pi^+\pi^-\pi$ decay as mediated by the $\rho$
resonance, $A_\rho (B\to \pi^+\pi^-\pi)$,
is the product of a $B\to\rho\pi$ amplitude and a $\rho\to\pi\pi$
vertex function, $\Gamma_{\rho\pi\pi}$. We give the relevant formulae
for $B\to\rho\pi \to \pi^+\pi^- \pi$ decay
in App.~\ref{app:Brhopi}.
In this section we focus on the construction
of $\Gamma_{\rho\pi\pi}$, generating a form which is consistent
with all known theoretical constraints.
We detail our procedure, as the form of $\Gamma_{\rho\pi\pi}$ is
important to the goals of the $B\to\rho\pi$ analysis: it drives
the size of the interference between $\rho$ states
produced in different regions of the Dalitz
plot. Our vertex function differs from that of
Ref.~\cite{dearho}, as the latter adopt a Breit-Wigner form.
We also compare our form with that adopted
in Ref.~\cite{babarbook}.
At this point we should mention that
the numerical differences are not large --- simply
because the pion vector form factor can be described fairly well by a
Breit-Wigner form. We also note
that a unitarized version of the vector form factor starting from
tree-level CHPT, including resonance fields, has been presented in
Ref.~\cite{ollerFV}, based on the methods described in
Sec.~\ref{sec:sff}. It could be used equally well as the
parametrization employed here.

The vector form factor $F_\rho(s)$ can be directly determined
from $e^+e^- \to \pi^+ \pi^-$ data in the $\rho$ resonance region.
General theoretical constraints guide its construction:
charge conservation requires $F_\rho(0)$ to be unity, and
time-reversal invariance and unitarity
lead to the identification of the phase of $F_\rho (s)$ with the
$l=1$, $I=1$ $\pi\pi$ phase shift, $\delta_1^1 (s)$, in the region
where $\pi\pi$ scattering is elastic, $s\lesssim (M_\pi + M_\omega)^2$.
Moreover, $F_\rho(s)$ is an analytic function in the complex $s$
plane, with a branch cut along the real axis beginning at the physical
threshold $s=4M_\pi^2$.
Below the two-pion cut at $s=4M_\pi^2$,
the vector form factor is real. Furthermore,
at small $s$ the form factor can be computed in CHPT,
as detailed in Refs.~\cite{GMII,BCT}.
All these constraints are
captured by the Muskhelishvili-Omn\`es (MO) integral
equation~\cite{MOmnes}. For $s\lesssim (M_\pi+M_\omega)^2$,
its solution can be written\cite{heyn}\footnote{The solution of
the MO equation with inelastic unitarity, important for
$s\gtrsim (M_\pi+M_\omega)^2$, has been discussed in Ref.~\cite{truongrho}.}
\begin{equation}
F_\rho (s) = P(s) \, \Omega(s)~,
\end{equation}
where $P(s)$ is a real polynomial and the Omn\`es function,
$\Omega(s)$, contains all the phase information,
\begin{eqnarray}
\Omega(s) &=& \exp \left( \frac{s}{\pi} \int_{4M_\pi^2}^\infty
\frac{ds}{s'} \frac{\phi_1 (s')}{s' - s - i \epsilon} \right)~,
\nonumber \\
\tan\phi_1(s) &\equiv& \frac{{\rm Im}\,F_\rho(s)}{{\rm Re}\,F_\rho(s)}
= \tan \delta_1^1(s)\,,
\end{eqnarray}
where $\delta_1^1(s)$ is the phase shift of $I=1$, $L=1$ scattering.
In the Heyn-Lang parametrization~\cite{heyn},
$\Omega(s)$ is approximated by the quotient of two analytic functions,
which contain polynomial pieces and the one-pion-loop expression for
the $\rho$ self-energy function.
$P(s)$ is chosen to be of third order in $s$ in Ref.~\cite{heyn}.
We use here a recent update of the pion form factor \cite{svghbo},
based on the Heyn-Lang parametrization.
Specifically, we use the parameter set of ``solution B''
of Ref.~\cite{svghbo}, reflecting a fit to the
$e^+e^-\to \pi^+\pi^-$ data in the elastic region, subject to the
constraint that the model reproduces the empirical $\pi\pi$
scattering length in the $I=1$, $L=1$ channel,
$a_1^1=(0.038\pm 0.0.002)\, {M_\pi}^{-3}$~\cite{nagels}.
In what follows, we neglect the presence of the $\omega$ resonance, or
effectively $\rho^0-\omega$ mixing. The latter
is an important isospin-violating effect visible in the
$e^+ e^- \to \pi^+ \pi^-$ data in the close vicinity of $s=M_\omega^2$
---  the fits of Ref.~\cite{svghbo} do include it.

To realize the vertex function $\Gamma_{\rho\pi\pi}(s)$, 
we define
\begin{equation}
\label{ourrho}
\Gamma_{\rho\pi\pi}(s)\equiv\frac{-F_\rho(s)}{f_{\rho\gamma}}\,,
\end{equation}
where, as described in App.~\ref{app:Brhopi}, the electromagnetic
coupling constant of the $\rho$ meson, $f_{\rho\gamma}$, is
$f_{\rho \gamma}=0.122\pm 0.001 \; \hbox{GeV}^2$~\cite{svghbo2}.
Ref.~\cite{dearho} adopts a Breit-Wigner form
for the vertex function, namely
\begin{equation}
\label{rhoffbw}
\Gamma_{\rho\pi\pi}^{\rm BW}(s)=
\frac{g_{\rho}}{ s - M_\rho^2 + i\Gamma_\rho M_\rho} \;,
\end{equation}
with the parameters $g_\rho=5.8$ and $\Gamma_\rho=150$ MeV
--- we use $M_\rho=769.3$ MeV\cite{pdg}.
The two forms are compared in Fig.~\ref{fig:rho} --- the Breit-Wigner
form offers a reasonable description of the vector form factor,
though differences can be seen.  In particular, the imaginary part
of the Breit-Wigner form does not vanish below physical threshold,
as it ought. This deficiency can be repaired by giving the
Breit-Wigner form a running width, i.e.,
\begin{eqnarray}
\label{rhoffBB}
\Gamma_{\rho\pi\pi}^{\rm RW}(s)
&=& \frac{g_{\rho}}{ s - M_\rho^2 + i\Pi(s)} \;, \\
\Pi(s) &=& \frac{M_\rho^2}{\sqrt{s}}\left(
\frac{p(s)}{p(M_\rho^2)} \right)^3 \Gamma_\rho\;,
\end{eqnarray}
where
$p(s)=\sqrt{s/4 - M_\pi^2}$.
This form, modulo the proportionality constant,
is adopted by Ref.~\cite{babarbook}. Fig.~\ref{fig:rhoBB}
compares Eqs.~(\ref{ourrho}) and (\ref{rhoffBB}) --- the two
forms are really very similar, though the forms differ slightly
as $s\to 4M_\pi^2$. The phase of the form factors, namely
$\tan^{-1}(({\rm Im\,} \Gamma_{\rho\pi\pi})/({\rm Re\,} \Gamma_{\rho\pi\pi}))$,
is plotted in Fig.~\ref{fig:phase}. Unitarity and time-reversal
invariance dictates that the phase be that of $I=1$, $L=1$ scattering;
the phase shifts from the data of Refs.~\protect{\cite{hyams,proto}}
are shown for comparison. The forms of Eq.~(\ref{ourrho}) and
Eq.~(\ref{rhoffBB}) confront the phase shift data nicely. The agreement
of the latter form is a particular surprise, as it contains
only two free parameters. Apparently the form of the
imaginary part is precisely
captured by the $\pi\pi$ branch cut, so that the phase is
accurately determined by an arctan prescription to unitarize the amplitude.
This is also realized in the CHPT
analysis including an explicit $\rho$
meson, for details see Ref.~\cite{BKMres}.

In the application to follow, we need to evaluate $\Gamma_{\rho\pi\pi}(s)$
for $s> (M_\pi+ M_\omega)^2$, so that at
$\sqrt{s} = (M_\pi + M_\omega) \simeq 920\,$MeV, the
form factor of Eq.~(\ref{ourrho}) is matched to the
Breit-Wigner form of Eq.~(\ref{rhoffbw}), yielding
${\rm Re}~F_\rho (s) \sim a'/s$ and ${\rm Im}~F_\rho (s)
\sim b'/s^2$ for large $s$. We now turn to a discussion of
our numerical results in $\rho$ and $\sigma$-mediated
$B\to 3\pi$ decay.

\medskip

\section{Results and discussion}
\label{sec:res}

First, we must collect parameters. For crisp comparison with
Refs.~\cite{deA,dearho}, we adopt the parameters used therein but
reiterate them here for convenience.
For meson masses and widths we use
$M_B = 5.279\,$GeV, $M_\pi = 139.57\,$MeV,
$M_\rho = 769.3\,$MeV, $\Gamma_\rho = 150\,$MeV,
$M_\sigma = 478\,$MeV, and $\Gamma_\sigma = 324\,$MeV.
We neglect the $B^+,\, B^0$ lifetime difference 
and use $\tau_B=1.6\cdot 10^{-12}\,$sec.
For quark masses, we use $m_b=4.6$ GeV and $\hat{m}=6$ MeV.
As for the CKM matrix elements,
we adopt the Wolfenstein parametrization~\cite{wolf83},
retaining terms of ${\cal O}(\lambda^3)$ in the real part and of
${\cal O}(\lambda^5)$ in the imaginary part, using 
$A = 0.806$, $\rho = 0.05$, $\eta = 0.36$, and $\lambda =0.2196$.
For the Wilson coefficients, we use $C_1=1.100$, $C_2=-0.226$,
$C_3=0.012$, $C_4=-0.029$, $C_5=0.009$, and $C_6=-0.033$, after
Ref.~\cite{buraswc}. For form factors and coupling constants we use
$F_0^{(B\to\sigma)}(M_\pi^2) = 0.46$, after Ref.~\cite{GNPT},
$F_1^{(B\to\pi)}(M_\rho^2) = 0.37$,
$A_0^{(B\to\rho)}(M_\pi^2) = 0.29$,
$f_\pi = \sqrt{2}(92.4\; \hbox{MeV})\simeq 131\,$MeV, 
and $f_\rho = 0.15\,$GeV$^2$.
Finally, we use $g_\rho=5.8$ and $g_{\sigma\pi\pi}=2.52$ GeV
when using the form factors of Refs.~\cite{deA,dearho}.

Let us begin by computing the
branching ratios
for $B\to\rho\pi$ and $B\to \sigma\pi$ decay. Assuming two-body
phase space, we use
Eqs.~(\ref{def2B0}) and (\ref{def2Bm}) with
Eqs.~(\ref{defetap}-\ref{defteta0}), as well as Eq.~(\ref{Btosigmapim}).
The results are tabulated in the first row of
Table \ref{tab:2body}. In the treatment
of Ref.~\cite{dearho},
the branching ratios of $B\to M_1 M_2$ decay
and its charge conjugate are identical,
even with penguin contributions, as no strong phase between
the amplitudes of differing weak phase has been included.

\renewcommand{\arraystretch}{1.2}
\begin{table}[hbt]
\begin{center}
\centerline{\parbox{15cm}
{\caption{Branching ratios (in units of $10^{-6}$) for
$B\to \rho\pi$ and $B\to \sigma\pi$ decay, computed at tree level.
The numbers in parentheses include
penguin contributions 
as well, after Ref.~\protect{\cite{dearho}}.
The first row of numbers compute the branching ratios using
two-body phase space. Although
${\rm Br}(\rho\to\pi^+\pi^-)\simeq 1$,
${\rm Br}(\sigma\to\pi^+\pi^-)\simeq 2/3$, so
that the numbers in brackets reflect the branching ratio
times $2/3$.
The rows labelled ``3-body'' compute
$B\to \rho\pi\to 3\pi$ and $B\to \sigma\pi\to \pi^+\pi^-\pi$ decay,
integrating over the entire three-body phase space.
``BW'' denotes the use of the form factors of
Refs.~\protect{\cite{deA,dearho}}, Eqs.~(\protect{\ref{GammadeA}},
\protect{\ref{rhoffbw}}), whereas
``RW'' denotes the use of the vector form factor of
Ref.~\protect{\cite{babarbook}}, Eq.~(\protect{\ref{rhoffBB}}).
Finally, ``$\ast$'' denotes the use of
the form factors we have advocated.
         \label{tab:2body}
}}}
\vspace{.3cm}
\begin{tabular}{|l||c|c|c|c||c|c|}
    \hline
     &  $\bar B^0 \to \rho^- \pi^+$
     &  $\bar B^0 \to \rho^+ \pi^-$
     &  $\bar B^0 \to \rho^0 \pi^0$
     &  $B^-\to \rho^0\pi^-$
     &  $B^-\to \sigma\pi^-$
     &  $\bar B^0 \to \sigma\pi^0$ \\ \hline
2-body   &  21.6 (21.0) & 5.96 (5.94) & 0.237 (0.308)
         &  4.74 (5.00) & 15.6 [10.4]       & 0.147 [0.0982] \\
  \hline
3-body (BW)  &  22.5 (22.1) & --- & ---
         &  4.11 (4.33) & 8.31        & 0.0739 \\
   \hline
3-body (RW)  &  22.4 (22.0) & --- & ---
         &  4.08 (4.30) & --- & --- \\
   \hline
3-body ($\ast$)  &  22.3 (21.9) & --- & ---
         &  4.03 (4.25) & 11.7        & 0.108 \\
   \hline
\end{tabular}
\end{center}
\end{table}

Proceeding to treat $B\to \rho\pi \to 3\pi$ and
$B\to \sigma\pi \to \pi^+\pi^-\pi$ decay,
we follow the $\rho$ and $\sigma$
intermediate states to their $\pi\pi$ final states.  We realize
the transition amplitudes as per Eqs.~(\ref{defArho0}), (\ref{defArhom}),
or (\ref{Asigpipi}), and integrate over the three-body phase
space as per Eq.~(\ref{phasespace}), computing the integral in $t$
over $[2M_\pi,M_B-M_\pi]$. With this procedure, the branching ratios for
$B\to\rho^-\pi^+$, $B\to\rho^+\pi^-$, and $B\to\rho^0\pi^0$ become
identical; we simply report the final result in the $B\to\rho^-\pi^+$
column. In treating $B^-\to\pi^+\pi^-\pi^-$ decay,
we divide the total rate by $1/2$, to compensate
for integrating over equivalent configurations.
Noting that ${\rm Br}(\sigma\to\pi^+\pi^-)\simeq 2/3$,
the quantity in brackets in the first row includes the factor
of $2/3$, for comparison with the three-body results.
Comparing the two-body branching ratios with those computed by
integrating over the entire three-body phase space,
it is evident that the branching ratios do not agree.
The deviations can be attributed to both
interference effects and finite-width effects.
As an example of the former, both the diagrams
illustrated in Figs.~\ref{fig:fb1} and \ref{fig:fb2}, as well as
those with $p_1\leftrightarrow p_2$, contribute to $B^-\to\pi^+\pi^-\pi^-$
decay. Clearly the interference of these diagrams is not included
when the $B^-\to\rho^0\pi^-$ or $B^-\to\sigma\pi^-$
process is treated as a two-body decay.
As an illustration of latter, note that
the couplings $g_\rho$ and $g_{\sigma\pi\pi}$ are typically
chosen so that they reproduce the $\rho\to\pi\pi$ and
$\sigma\to\pi^+\pi^-$ decay rates, namely
\begin{equation}
\label{coupsig}
\frac{2}{3} \Gamma_\sigma=\Gamma(\sigma\to\pi^+\pi^-) =
\frac{1}{16\pi M_\sigma^2}(M_\sigma^2 - 4 M_\pi^2)^{1/2} |g_{\sigma\pi\pi}|^2
\end{equation}
and
\begin{equation}
\label{couprho}
\Gamma_\rho=
\frac{1}{48\pi M_\rho^2}(M_\rho^2 - 4 M_\pi^2)^{1/2} |g_\rho|^2 \,.
\end{equation}
For the meson masses and widths we have used, these formulae yield
$g_{\sigma\pi\pi}=2.53$ and $g_\rho=6.03$, respectively.
Adopting these couplings in place of those used
in Refs.~\cite{deA,dearho} does reduce the
discrepancy. Note that it is a ``finite width'' effect in that reducing the
numerical width of the $\rho$ or $\sigma$ meson, in concert with
Eqs.~(\ref{coupsig},\ref{couprho}), reduces the discrepancy between
the two- and three-body treatments.
It is worth noting, however, for the physical values of the
meson widths, that there is no
one fixed coupling $g_{\sigma\pi\pi}$ or $g_\rho$ which
removes the discrepancy entirely --- the needed coupling
in any given case is sensitive to the form factor chosen,
as well as to the masses of the other particles in
the final state.
The former is apparent from a comparison with
the vector form factor of
Ref.~\cite{babarbook}, Eq.~(\ref{rhoffBB}), for
which we use $g_\rho=5.8$ as well. The normalization issue of which
we speak is particularly relevant for the comparison of theoretical
branching ratios, for $B\to VP$ decay, e.g., to experiment. It is
present regardless of the form factor used. That is, in the case of
the vector form factor we adopt, Eq.~(\ref{ourrho}), the determination
of $f_{\rho\gamma}$ can also be modified by finite width effects.
The sign and size of the  mismatch
between the two- and three-body phase space calculations can be
quite sensitive to the choice of form factor, as illustrated by
the scalar case. 
We set the normalization of the form factor of Sec.~\ref{sec:sff},
denoted by ``$\ast$'' in the Tables,
to that of the form factor of Ref.~\cite{deA}, Eq.~(\ref{GammadeA}).
This is the only manner in which the parameters $M_\sigma,\Gamma_\sigma$
enter our analysis. Were we to determine
the normalization so that the two- and three-body
computations of the $\bar B^0 \to \sigma \pi^0 \to \pi^+\pi^-\pi^0$
branching ratio yield identical results,
the effective impact of the $\sigma$ in the
$\rho\pi$ phase space would be reduced by some 10\%.

\renewcommand{\arraystretch}{1.2}
\begin{table}[hbt]
\begin{center}
\centerline{\parbox{15cm}
{\caption{Effective branching ratios (in units of $10^{-6}$) for
$B\to \rho\pi$ decay, computed at tree level.
The numbers in parentheses include
penguin contributions as well, after Ref.~\protect{\cite{dearho}}.
The form factors are defined as in Table \protect{\ref{tab:2body}}.
                   \label{tab:3body}}}}
\vspace{.3cm}
\begin{tabular}{||l||c|c|c|c||c||}
    \hline
$\delta$ [MeV] (f.f.)
     &  $\bar B^0 \to \rho^- \pi^+$
     &  $\bar B^0 \to \rho^+ \pi^-$
     &  $\bar B^0 \to \rho^0 \pi^0$
     &  $B^-\to \rho^0\pi^-$   & ${\cal R}$
\\ \hline \hline
200 (BW) & 15.1 (14.7) & 4.21 (4.24) & 0.508 (0.497) & 3.50 (3.68)
& 5.5 (5.1)\\
  \hline
300 (BW)  &  16.4 (16.0) & 4.74 (4.76) & 0.918 (0.908) & 3.89 (4.10)
& 5.4 (5.1) \\
   \hline
200 (RW)  & 15.1 (14.8) & 4.19 (4.21) & 0.468 (0.463)  &  3.49 (3.68)
& 5.5 (5.2) \\
   \hline
300 (RW)  &  16.4 (16.0) & 4.69 (4.70) & 0.835 (0.831) &  3.87 (4.07)
& 5.5 (5.1)  \\
   \hline
200 ($\ast$)   &  15.3 (14.9) & 4.26 (4.28) & 0.473 (0.467) &  3.49 (3.68)
& 5.6 (5.2)  \\
   \hline
300 ($\ast$)   & 16.4 (16.0)  & 4.75 (4.76) & 0.865 (0.859) &  3.85 (4.06)
& 5.5 (5.1)  \\
   \hline
   \hline
   \hline
$\delta$ [MeV] (f.f.)     &  $B^0 \to \rho^+ \pi^-$
     &  $B^0 \to \rho^- \pi^+$
     &  $B^0 \to \rho^0 \pi^0$
     &  $B^+\to \rho^0\pi^+$ & ${\bar{ \cal R}}$  \\
\hline
\hline
200 (BW)   &  15.1 (14.7) & 4.21 (4.15) & 0.508 (0.615)
         &  3.50 (3.68)
& 5.5 (5.1) \\
  \hline
300 (BW)  &  16.4 (16.0) & 4.74 (4.67) & 0.918 (1.02)
         &  3.89 (4.10)
& 5.4 (5.0) \\
   \hline
200 (RW)  &  15.1 (14.7) & 4.19 (4.13) & 0.468 (0.571)
         &  3.49 (3.68)
& 5.5 (5.2) \\
   \hline
300 (RW)  &  16.4 (15.9) & 4.69 (4.62) & 0.835 (0.935)
         &  3.87 (4.07)
& 5.5 (5.0) \\
   \hline
200 ($\ast$)  &  15.3 (14.8) & 4.26 (4.20) & 0.473 (0.576)
         &  3.49 (3.68)
& 5.6 (5.2) \\
   \hline
300 ($\ast$)   & 16.4 (15.9)  & 4.75 (4.68) & 0.865 (0.963)
         &  3.85 (4.06)
& 5.5 (5.1) \\
   \hline
   \hline
\end{tabular}
\end{center}
\end{table}

In Table \ref{tab:3body} we report the $B\to\rho\pi\to 3\pi$
branching ratios, computed in the manner of Ref.~\cite{dearho}.
Our numerical results differ slightly from theirs (note that
$B^0 \leftrightarrow \bar B^0$ in their Table III).
The upshot is that our estimate of
${\cal R}$ with penguin contributions is $\sim 5.1$, rather
than the $5.5$ they estimate. We show the branching ratios
computed for differing vector form factors; these differing
choices have little impact on the resulting branching ratios,
or on ${\cal R}$.

We compute the $B\to\sigma\pi$ branching ratios in
Table \ref{tab:cf}. In this case our computed branching
ratios, for $B^-\to\sigma\pi^-$ decay, are a factor of two larger
with the same form factors and parameters input,
as our formula, Eq.~(\ref{cfdeasig}), differs from theirs by a
factor of $\sqrt{2}$. Thus the impact of the $\sigma$ in the $\rho^0\pi^-$
phase space is rather larger than that estimated in Ref.~\cite{deA}.
Updating the scalar form factor to use what we feel is
its best
estimate, we find that the values of ${\cal R}$ are smaller still.
Interestingly, the computed values of ${\cal R}$ are comparable
to the empirical results, albeit the errors are large.
[An additional contribution to the phenomenological value of
${\cal R}$, realized through
a diagram mediated by the $a_1^-$ meson,
is proposed in Ref.~\cite{paver}.]

Turning to $B\to \sigma\pi^0$ decay, we see that the contribution
of the $\sigma$ meson to $B^0 (\bar B^0) \to \rho\pi$ decay
is {\it much} smaller ---
with the scalar form factor we advocate, the effect is some
10\%. Interestingly the $\sigma$ has a tremendous impact on
$B^-\to\rho^0\pi^-$ decay, and a relatively modest one on
$\bar B^0\to\rho^0\pi^0$ decay.
Let us emphasize
that we have realized our numerical analysis at tree level.
It is the relative size
of the penguin contributions
in $\bar B^0\to\sigma\pi^0$
and $\bar B^0\to\rho^0\pi^0$ decay
which is of relevance to the isospin analysis to extract $\alpha$.
The presence of the $\sigma\pi^0$ final state in the $\rho^0\pi^0$
phase space can
break the assumed relationship, Eq.~(\ref{isopenguin}), between
the penguin contributions in $\rho\pi$ and thus
mimic the effect of isospin violation --- alternatively we can
expand the $\rho\pi$ analysis to include the $\sigma\pi$ channel.
Nevertheless, we expect our estimates to be crudely indicative
of the importance of these effects --- quantitatively, however,
differences may exist. It is worth noting that the 
$\sigma\pi^0$ and $\rho^0\pi^0$ contributions can, to some measure,
be distinguished. 
Certainly the $\sigma\pi^0$ and $\rho\pi^0$ contributions behave
differently under the cut on the invariant mass of the $\pi^+\pi^-$
pair, recalling Eq.~(\ref{phasespace}). Moreover,
making a cut on the helicity angle $\theta$, defined in Eq.~(\ref{deftheta}),
ought also be helpful in separating the $\rho^0$ and $\sigma$
contributions.
This is illustrated in Figs.~\ref{fig:theta} and \ref{fig:theta0}.
The $\rho^0\pi$ contributions roughly follow a $\cos^2(\theta)$ distribution, 
whereas the $\sigma\pi$ contributions are quite flat, save for
the bump resulting from the $\Gamma_{\sigma\pi\pi}(u)$ term
in Eq.~(\ref{cfdeasig}). 
Cutting on the helicity angle $\theta$ should also
help disentangle the contributions from some of the $B^*$ resonances,
discussed in Ref.~\cite{dearho}. The contributions of $B^*$
resonances to the $\rho\pi$ channels should be
included in a more refined analysis, but they will not alter the
conclusions drawn here.

\renewcommand{\arraystretch}{1.2}
\begin{table}[hbt]
\begin{center}
\centerline{\parbox{15cm}
{\caption{Effective branching ratios (in units of $10^{-6}$) for
$B\to \sigma\pi$ and $B\to \rho\pi$
decay, computed at tree level.
The form factors are defined as in Table \protect{\ref{tab:2body}}.
                   \label{tab:cf}}}}
\vspace{.3cm}
\begin{tabular}{||l||c|c|c|c||c||}
    \hline
$\delta$ [MeV] (f.f.)
& $B^-\to \sigma\pi^-$ & $B^-\to (\rho^0 + \sigma)\pi^-$
& $\bar B^0 \to \sigma\pi^0$ & $\bar B^0 \to (\rho^0 + \sigma)\pi^0$ &
${\cal R}$
\\ \hline \hline
200 (BW) & 2.97 & 6.16 & 0.0258 & 0.516 &  3.1 \\
  \hline
300 (BW)  &  5.17 & 8.61 & 0.0457 & 0.940 & 2.5 \\
   \hline
200 (RW) & 2.97 & 6.19 & 0.0258 & 0.475 & 3.1 \\
  \hline
300 (RW)  &  5.17 & 8.62 & 0.0457 & 0.855 & 2.4 \\
   \hline
200 ($\ast$) & 4.11 & 7.61 & 0.0396 & 0.508 &  2.6 \\
  \hline
300 ($\ast$) &  7.01 & 10.7 & 0.0663 & 0.916 & 2.0 \\
   \hline
   \hline
\end{tabular}
\end{center}
\end{table}

\section{Summary}
\label{sec:sum}

In this paper, we have scrutinized the role of the $\sigma$ meson
in $B\to\rho\pi\to 3\pi$ decay,
understanding its dynamical origin in the strong pion-pion final
state interactions in the scalar-isoscalar channel.
The presence of the $\sigma\pi^0$ contribution
in the $\rho^0\pi^0$ phase space is important in that it can
break the assumed relationship between the penguin amplitudes,
Eq.~(\ref{isopenguin}), consequent to an assumption of isospin
symmetry. In this, then, its presence mimics the effect of isospin
violation. The
salient results of our investigation can be summarized as follows:

\begin{itemize}

\item[i)] We have considered how SM isospin violation can impact
 the analysis to extract $\alpha$ in
  $B\to\rho\pi$ decay. Under the assumption that
  $|\Delta I| = 3/2$ and $|\Delta I| = 5/2$ amplitudes share the
  same weak phase, the presence of an additional amplitude of
  $|\Delta I|=5/2$ character, induced by isospin-violating effects,
  does not impact the $B\to\rho\pi$ analysis in any way. This is
  in contradistinction to the isospin analysis in $B\to\pi\pi$.
  Thus the isospin-violating effects of importance are those which
  can break the assumed relationship between the penguin contributions,
  Eq.~(\ref{isopenguin}).

\item[ii)]
  The scalar form factor
can be determined to good precision by
combining the constraints of chiral symmetry,
analyticity, and unitarity. The form factor we adopt describes
the appearance of the $f_0(980)$ as well, so that the shape of the
$f_0(980)$ contribution in $B\to f_0(980)\pi\to 3\pi$, e.g., should serve
 as a test of our approach. We emphasize that
the resulting
scalar form factor is very different from the commonly used Breit-Wigner
  form with a running width.  This is in stark contrast
  to the vector form factor, which is dominated by the $\rho$ resonance.
  In that case, one can construct  simple forms that fit the
  theoretical and empirical constraints.

\item[iii)] We have pointed out that the two- versus three-body
  treatments of the decays $B \to \rho \pi, B\to \sigma \pi$
  can lead to differing results due to finite-width and interference
  effects. 

\item[iv)] Remarkably, the impact of the $\sigma\pi$ channel on the
  ratio ${\cal R}$, cf. Eq.~(\ref{rhoratio}), is huge. The numbers
  we find for ${\cal R}$
  are in agreement with the empirical ones, given its sizeable
  experimental uncertainty. This underscores the suggestion made, as well
  as improves the calculations done, in Ref.~\cite{deA}. Our
  analysis is based on 
  {\em consistent} scalar and vector form factors.

\item[v)] On the other hand, the impact of the $\sigma \pi$
  channel  on the $B\to\rho\pi$ isospin analysis is merely significant.
  Varying the cuts on the $\pi\pi$ invariant mass and helicity angle
  $\theta$ should be helpful in disentangling the various contributions.

\item[vi)] We have shown that one can expand the isospin analysis
  to include the $\sigma\pi$ channel because it has definite
  properties under CP. This may be necessary if varying the
  cuts in the $\pi\pi$ invariant mass and helicity angle $\theta$
  are not sufficiently effective in suppressing the contribution
  from the $\sigma\pi^0$ channel in the $\rho^0\pi^0$ phase space.

\end{itemize}

\noindent
This work is merely a first step in
exploiting constraints from chiral symmetry, analyticity, and
unitarity in the description of hadronic B  decays.
In particular, the contribution
of the ``doubly''  OZI-violating strange scalar form factor
and its phenomenological role in factorization breaking
ought be investigated.

\medskip
\noindent{\bf Acknowledgements}
We thank H.R. Quinn for helpful discussions and
M. Gronau for remarks concerning rescattering effects in B decays.
We thank J.A. Oller and J. Tandean for collaborative discussions on topics
germane to the issues raised here. In particular,
we gratefully acknowledge J.A. Oller for the use of
his scalar form factor program.
S.G. thanks the SLAC Theory Group and the
Aspen Center for Physics for hospitality and is supported by
the U.S. Department of Energy under contract DE-FG02-96ER40989.

%%%%%%%%%%%%% appendix A: rho contribution %%%%%%%%%%%%%%%%%
\appendix
\def\theequation{\Alph{section}.\arabic{equation}}
\setcounter{equation}{0}
\section{Formulae for \boldmath{$B\to \rho \pi\to \pi^+\pi^-\pi^0$} decay}
\label{app:Brhopi}

\begin{figure}[htb]
\begin{center}
\epsfig{file=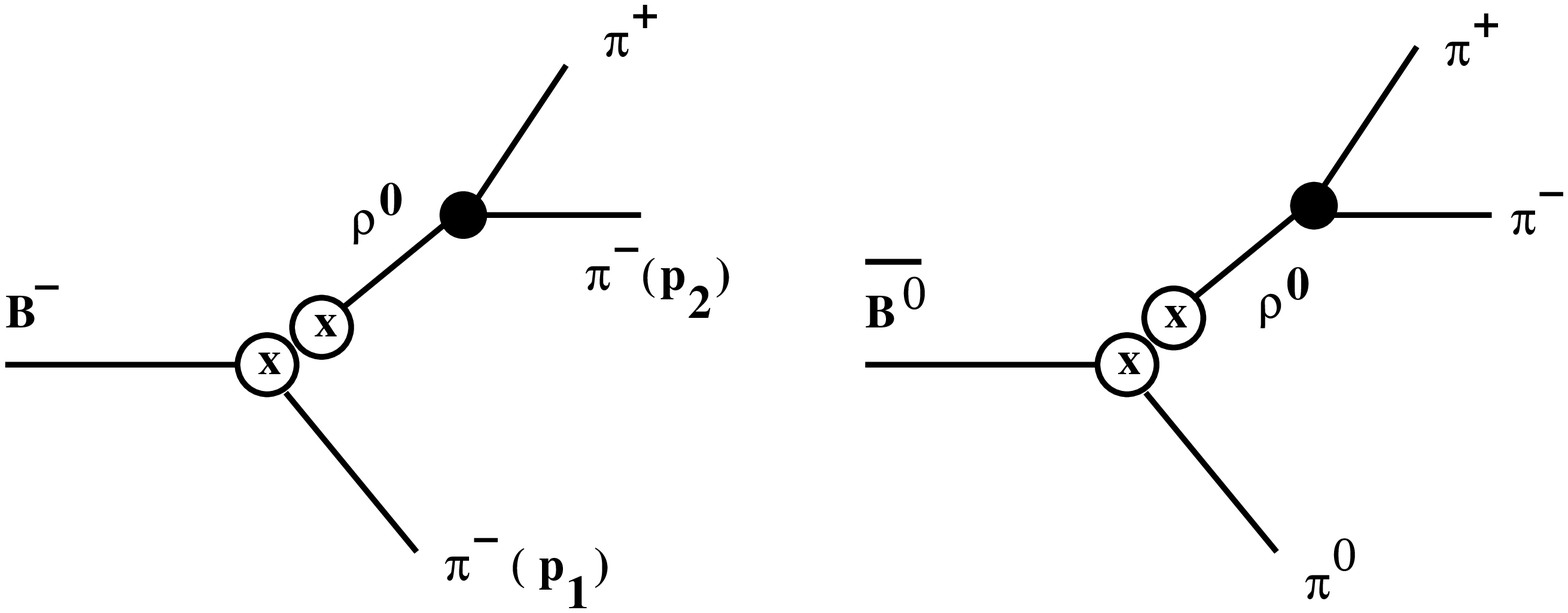, height = 5cm}
\end{center}
\centerline{\parbox{12cm}{\caption{
$B\to \pi^+\pi^-\pi$ decay as mediated by the $\rho$ resonance.
The factorized weak vertex is denoted by ``$\otimes\otimes$''.
The filled circle denotes the strong three-meson
vertex, here $\rho \to 2\pi$.
\label{fig:fb2}
}}}
\end{figure}

\begin{figure}[htb]
\begin{center}
\epsfig{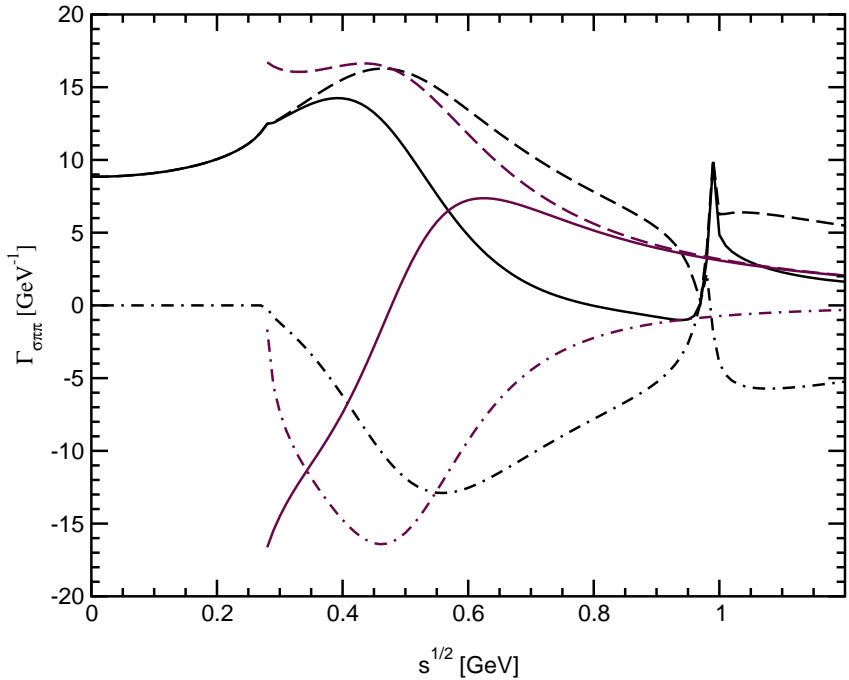}
\end{center}
\centerline{\parbox{12cm}{\caption{
The $\sigma\to\pi^+(p_+)\pi^-(p_-)$ form factor $\Gamma_{\sigma\pi\pi}$ as
a function of $\sqrt{s}$, with $s=(p_+ + p_-)^2$.
The real (solid line) and imaginary (dot-dashed line) parts of
$\Gamma_{\sigma\pi\pi}$, as well as its modulus (dashed line), are
shown. The curves which do not persist below physical threshold,
$\sqrt{s}=2M_{\pi} \sim 0.27$ GeV, correspond to the form factor
adopted in Ref.~\protect\cite{deA}, whereas the curves which extend to
$s=0$ correspond to the form factor adopted here \cite{OM}.
\label{fig:sff}
}}}
\end{figure}

\begin{figure}[htb]
\begin{center}
\epsfig{file=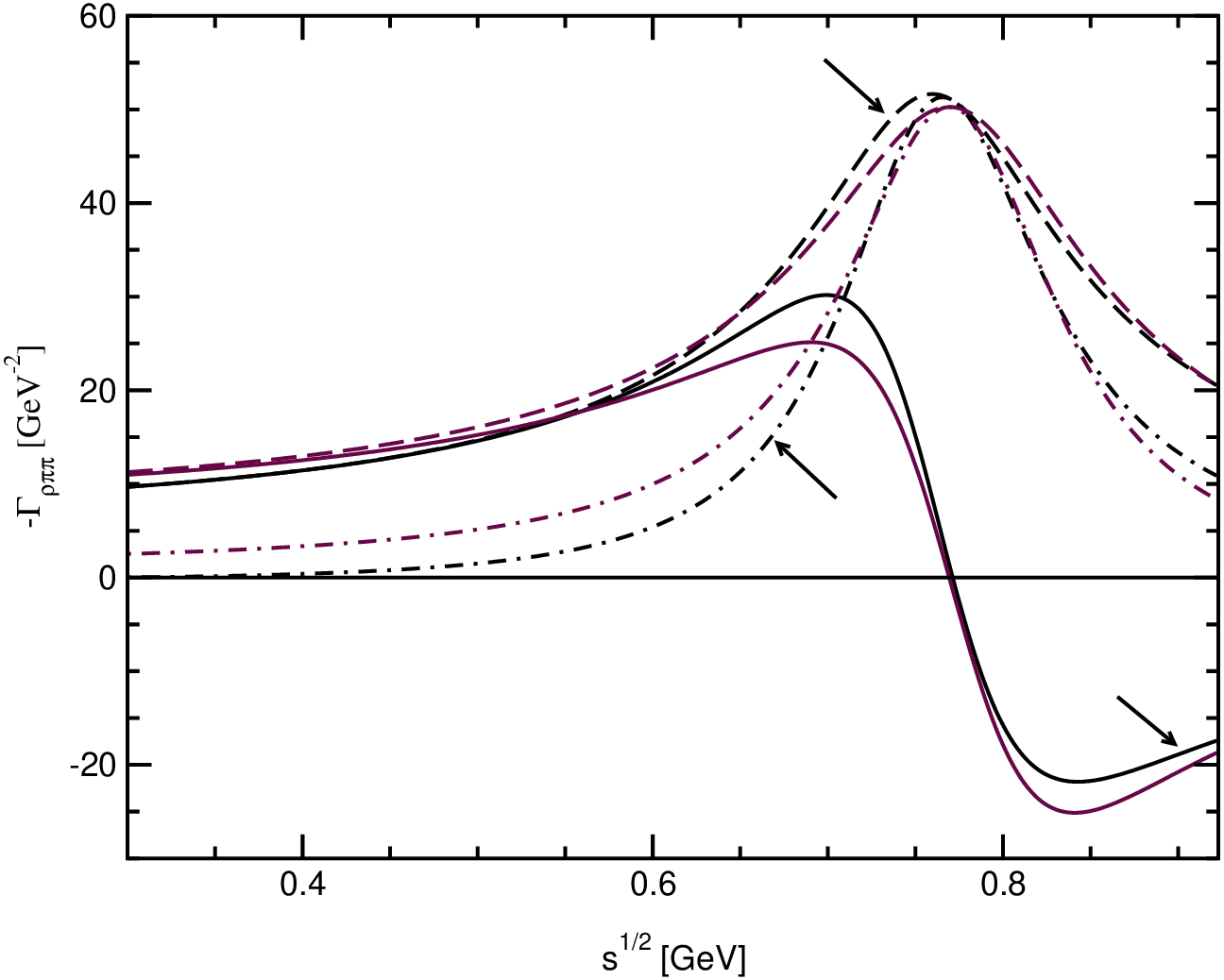, height = 10cm}
\end{center}
\centerline{\parbox{12cm}{\caption{
The $\rho\to\pi^+(p_+)\pi^-(p_-)$ form factor $-\Gamma_{\rho\pi\pi}$ as
a function of $\sqrt{s}$, with $s=(p_+ + p_-)^2$. The form factor
is shown in the region for which $l=1$, $I=1$ $\pi\pi$ scattering
is elastic.
The real (solid line) and imaginary (dot-dashed line) parts of
$\Gamma_{\rho\pi\pi}$, as well as its modulus (dashed line), are
shown. Noting Eq.~(\protect{\ref{ourrho}}), the arrows indicate
the form factor given by
$F_\rho(s)/f_{\rho\gamma}$, as detailed in Sec.~\protect{\ref{sec:vector}},
whereas the other curves correspond
to the Breit-Wigner form $-g_\rho/(s - M_\rho^2 + i M_\rho \Gamma_\rho)$,
adopted in Ref.~\protect{\cite{dearho}}.
\label{fig:rho}
}}}
\end{figure}

\begin{figure}[htb]
\begin{center}
\epsfig{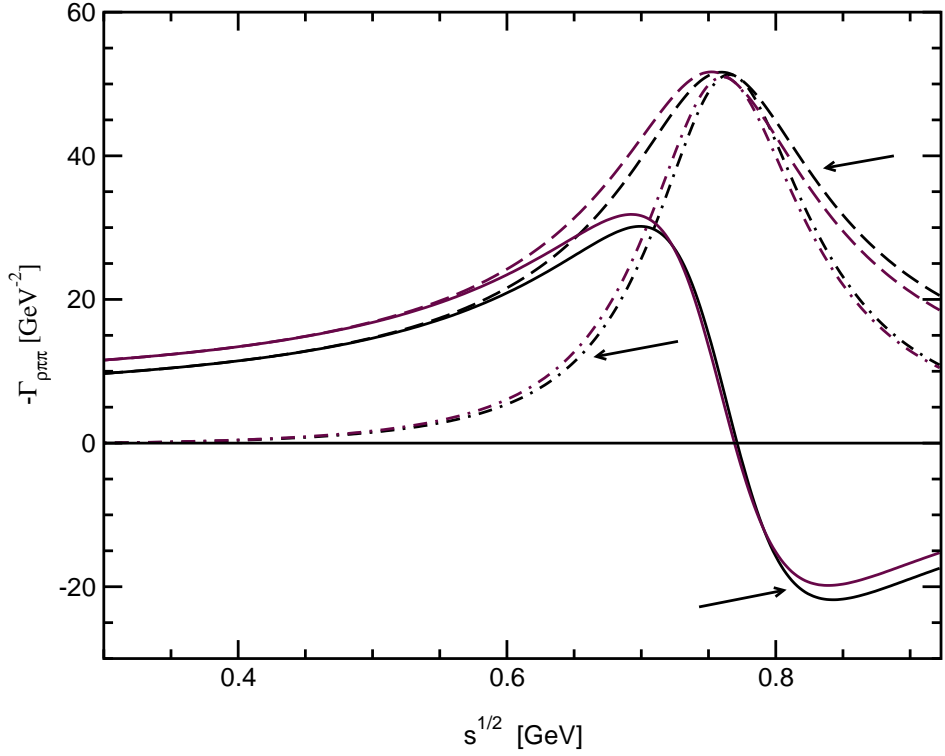}
\end{center}
\centerline{\parbox{12cm}{\caption{
The $\rho\to\pi^+(p_+)\pi^-(p_-)$ form factor $-\Gamma_{\rho\pi\pi}$ as
a function of $\sqrt{s}$, with $s=(p_+ + p_-)^2$. The form factor
is shown in the region for which $l=1$, $I=1$ $\pi\pi$ scattering
is elastic.
The real (solid line) and imaginary (dot-dashed line) parts of
$\Gamma_{\rho\pi\pi}$, as well as its modulus (dashed line), are
shown. Noting Eq.~(\ref{ourrho}), the arrows indicate
the form factor given by
$F_\rho(s)/f_{\rho\gamma}$,
whereas the other curves
correspond to the form of Eq.~(\protect{\ref{rhoffBB}})
adopted in Ref.~\protect{\cite{babarbook}}.
\label{fig:rhoBB}
}}}
\end{figure}

\begin{figure}[htb]
\begin{center}
\epsfig{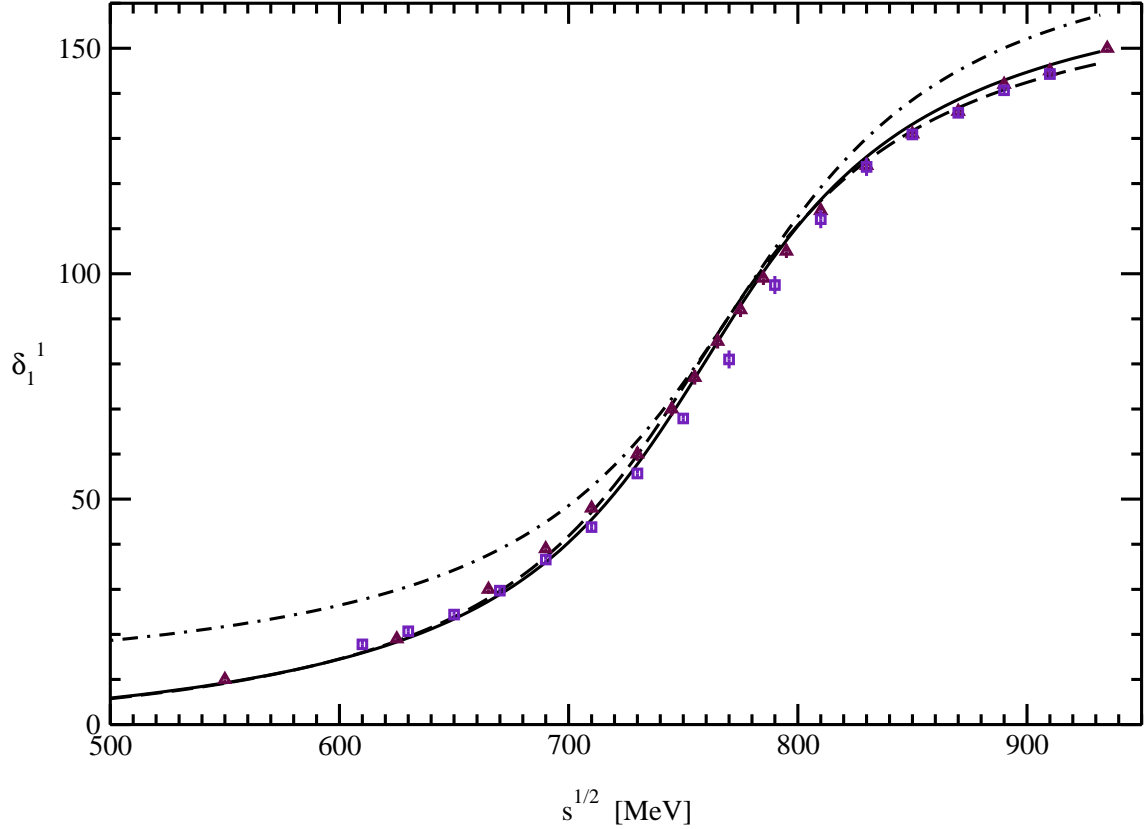}
\end{center}
\centerline{\parbox{12cm}{\caption{
The phase of the vector
form factor $\Gamma_{\rho\pi\pi}(s)$ as
a function of $\sqrt{s}$, 
in the region where the scattering is elastic.
The form factor we adopt, $F_\rho(s)/f_{\rho\gamma}$ (solid line),
the relativistic
Breit-Wigner form of Ref.~\protect{\cite{dearho}} (dot-dashed line),
as well as that of Ref.~\protect{\cite{babarbook}} (dashed line),
are all shown. Unitarity and time-reversal invariance requires that the phase
be the phase shift $\delta_1^1$ of $I=1$, $L=1$ $\pi$-$\pi$
scattering. The empirical phase shifts of
Ref.~\protect{\cite{hyams}}
($\square$) and Ref.~\protect{\cite{proto}}
($\vartriangle$) are indicated.
\label{fig:phase}
}}}
\end{figure}

\begin{figure}[htb]
\begin{center}
\epsfig{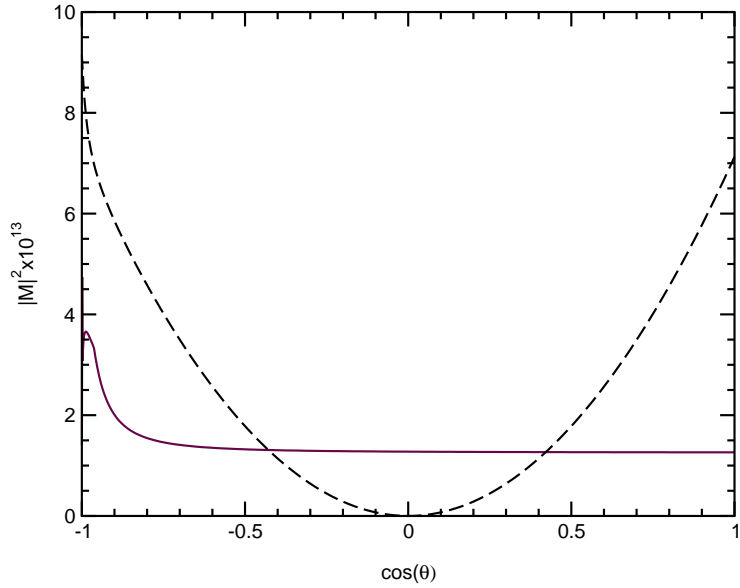}
\end{center}
\centerline{\parbox{12cm}{\caption{
Absolute square of the matrix element,
$|M|^2$, for $B^-\to\rho^0\pi^-$ decay (dashed line),
Eq.~(\protect{\ref{defArhom}}),
and for $B^-\to\sigma\pi^-$ decay (solid line),
Eq.~(\protect{\ref{cfdeasig}}), as a function of
$\cos\theta$ at $t=M_\rho^2$. 
The scalar and vector form factors advocated in Secs.
\protect{\ref{sec:sff}} and \protect{\ref{sec:vector}}
have been used. 
The bump in the solid line
reflects the presence of the $\Gamma_{\sigma\pi\pi}(u)$ term
in Eq.~(\protect{\ref{cfdeasig}}).
\label{fig:theta}
}}}
\end{figure}

\begin{figure}[htb]
\begin{center}
\epsfig{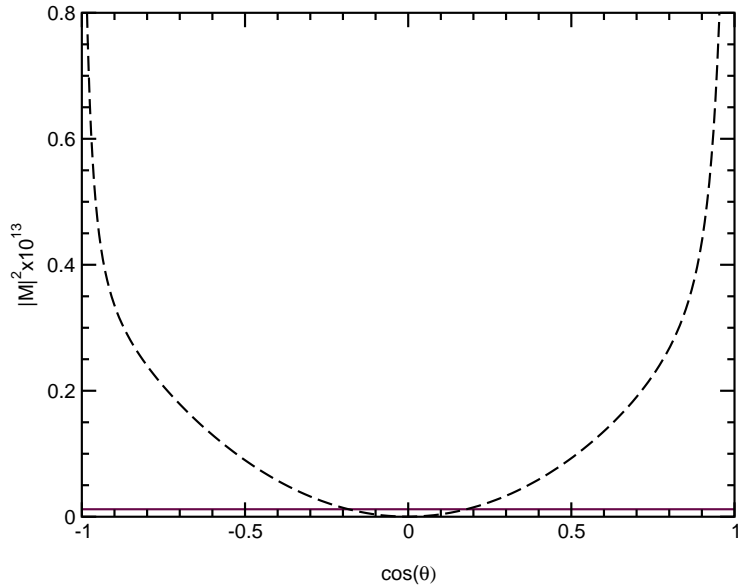}
\end{center}
\centerline{\parbox{12cm}{\caption{
Absolute square of the matrix element,
$|M|^2$, for $\bar B^0\to\rho^0\pi^0$ decay (dashed line),
Eq.~(\protect{\ref{defArho0}}),
and for $B^0\to\sigma\pi^0$ decay (solid line),
Eq.~(\protect{\ref{cfdeasig0}}), as a function of
$\cos\theta$ at $t=M_\rho^2$.
The scalar and vector form factors advocated in Secs.
\protect{\ref{sec:sff}} and \protect{\ref{sec:vector}}
have been used. 
\label{fig:theta0}
}}}
\end{figure}

In this appendix, we report the formulae needed to evaluate
$B\to \rho \pi\to \pi^+\pi^-\pi^0$ transitions, as per the
approach of Sec.~\ref{sec:evalB3pi}. For clarity of comparison,
we conform as much as possible to the notation and conventions of
Ref.~\cite{dearho}, but
give the formulae required
for completeness. This also allows us to identify the
changes in replacing the Breit-Wigner form adopted for
the $\rho$ resonance in Ref.~\cite{dearho}
with the pion vector form factor we
discuss in Sec.~\ref{sec:vector}. Defining
\begin{equation}
\label{grho2pi}
\langle \pi^0 (p_2) \pi^- (p_1)
| \rho^- (p_\rho, \epsilon) \rangle
= g_\rho \, {\epsilon} \cdot (p_2 - p_1)
\end{equation}
and
\begin{eqnarray}
\label{def2B0}
\langle \rho^i(p_\rho) \pi(p_\pi) |
{\cal H}_{\rm eff} | \bar B^0(p_B) \rangle
&=& 2\epsilon^\ast\cdot p_{\pi} \eta^i \;, \\
\label{def2Bm}
\langle \rho^0(p_\rho) \pi^-(p_\pi) |
{\cal H}_{\rm eff} | B^-(p_B) \rangle
&=& 2\epsilon^\ast\cdot p_{\pi} \tilde \eta^0 \;,
\end{eqnarray}
where $i\in (+,0,-)$,
the $\bar B^0\to \rho\pi \to \pi^+\pi^-\pi^0$ amplitude can be written
\begin{equation}
\label{defArho0}
A_\rho(\bar B^0(p_B) \to \pi^+(p_+)\pi^-(p_1)\pi^0(p_2))
= -\eta^0 (s-u)\Gamma_{\rho\pi\pi}(t)  + \eta^+ (s-t) \Gamma_{\rho\pi\pi}(u)
+ \eta^- (t-u) \Gamma_{\rho\pi\pi}(s)\;,
\end{equation}
where the first of these contributions is illustrated in Fig.~\ref{fig:fb2}.
We have used $s=(p_1 + p_2)^2$, $t=(p_+ + p_1)^2$, and $u=(p_+ + p_2)^2$,
and have summed over the polarization states of the $\rho^i$ mesons,
setting $M_{\pi^\pm}=M_{\pi^0}$. With our conventions for the flavor
content of the meson states, we see that 
$|{\pi^+}\rangle \,, |{\rho^+}\rangle\, =-|{I=1\,I_3=1}\rangle$, 
whereas the other $\pi$ and $\rho$ charge states do not have
a minus sign when written in the isospin 
basis. Using the isospin-raising operator $\tau_+$,
we thus determine from Eq.~(\ref{grho2pi}) that
$\langle \pi^+ (p_+) \pi^- (p_1)
| \rho^0 (p_\rho, \epsilon) \rangle
= -g_\rho \, {\epsilon} \cdot (p_+ - p_1)$ and
$\langle \pi^+ (p_+) \pi^0 (p_2)
| \rho^+ (p_\rho, \epsilon) \rangle
= g_\rho \, {\epsilon} \cdot (p_+ - p_2)$; the signs we indicate
consequently 
follow\footnote{We thank J. Tandean for discussions on this point.}.
For the $B^-\to \rho^0\pi^- \to \pi^+\pi^-\pi^-$ amplitude we have
\begin{equation}
\label{defArhom}
A_\rho(B^-(p_B) \to \pi^+(p_+)\pi^-(p_1)\pi^-(p_2))
= - \tilde \eta^0 \left[
(s-u)\Gamma_{\rho\pi\pi}(t)  + (s-t) \Gamma_{\rho\pi\pi}(u) \right] \;.
\end{equation}
Note that $\Gamma_{\rho\pi\pi}(s)$
is the pion vector form factor, for which a Breit-Wigner form
is used in Ref.~\cite{deA}.
As discussed in Sec.~\ref{sec:vector}, we
replace
\begin{equation}
\label{defGrhopipi}
\Gamma_{\rho\pi\pi}(x)=
\frac{g_\rho}{ x - M_\rho^2 + i\Gamma_\rho M_\rho} \to
\frac{-F_\rho(x)}{f_{\rho\gamma}}~,
\end{equation}
where $f_{\rho\gamma}$ is the
electromagnetic
coupling constant of the $\rho$ meson, determined from
\begin{equation}
\Gamma(\rho\to e^+ e^-)= \frac{4\pi \alpha^2}{3 M_\rho^3}f_{\rho\gamma}^2 \;,
\end{equation}
where $\Gamma(\rho\to e^+ e^-)$ is, in turn, extracted from
$e^+ e^-\to \pi^+\pi^-$ data at $s=M_\rho^2$,
as described in Ref.~\cite{svghbo2}.
For the ``solution B'' fit of Ref.~\cite{svghbo} we have
$f_{\rho \gamma}=0.122\pm 0.001 \; \hbox{GeV}^2$~\cite{svghbo2}.
The two forms are compared in Fig.~\ref{fig:rho}. In Ref.~\cite{dearho}
the parameters $g_\rho=5.8$ and $\Gamma_\rho=150$ MeV are chosen --- we 
use the value $M_\rho=769.3$ MeV\cite{pdg} for the $\rho$ meson mass, as
it is not reported in Ref.~\cite{dearho}.

To determine $\eta^i$, $\bar\eta^0$, we introduce
\begin{eqnarray}
\langle \rho^-(p_\rho,\epsilon) | \bar d \gamma_\mu u | 0 \rangle
&=&f_{\rho} \epsilon^{\ast}_{\mu}\;, \\
q^\mu \langle \rho^-(p_\rho,\epsilon) | \bar u \gamma_\mu
(1-\gamma_5) b | \bar B^0(p_B) \rangle&=& -i\, 2M_\rho
(\epsilon^\ast \cdot q)A_0^{B\to\rho}(q^2)\;,
\end{eqnarray}
where $q=p_B - p_\rho$,
and recall Eqs.~(\ref{fpidef}) and (\ref{Btopidef}), to find
\begin{eqnarray}
\label{defetap}
\!\!\!\!\!\!\!&&\eta^+ = \frac{G_F}{\sqrt{2}} 
\left[ \lambda_u a_1 - \lambda_t a_4 +
\lambda_t \frac{a_6 M_\pi^2}{\hat{m}(m_b+\hat{m})} \right]
f_\pi M_\rho A_0^{B\to\rho}(M_\pi^2) \;,\\
\label{defetam}
%!
\!\!\!\!\!\!\!&&\eta^- =
\frac{G_F}{\sqrt{2}} 
\left[ \lambda_u a_1 - \lambda_t a_4 \right]
f_\rho F_1^{B\to\pi}(M_\rho^2) \;,\\
\label{defeta0}
\!\!\!\!\!\!\!&&\eta^0 =
-\frac{G_F}{2\sqrt{2}}
\Bigg\{\!\!
\left[ \lambda_u a_2 + \lambda_t a_4 -
\lambda_t \frac{a_6 M_\pi^2}{\hat{m}(m_b+\hat{m})} \right]
f_\pi M_\rho A_0^{B\to\rho}(M_\pi^2)
+ \left[ \lambda_u a_2 + \lambda_t a_4\right] f_\rho F_1^{B\to\pi}(M_\rho^2)
\!\Bigg\} \;,\\
\label{defteta0}
\!\!\!\!\!\!\!&&\tilde\eta^0 =
\frac{G_F}{2} 
\Bigg\{\!\!
\left[ \lambda_u a_1 - \lambda_t a_4 +
\lambda_t \frac{a_6 M_\pi^2}{\hat{m}(m_b+\hat{m})} \right]
f_\pi M_\rho A_0^{B\to\rho}(M_\pi^2)
+ \left[ \lambda_u a_2 + \lambda_t a_4\right] f_\rho F_1^{B\to\pi}(M_\rho^2)
\!\Bigg\} \;,
\end{eqnarray}
where we
neglect electroweak penguin contributions,
as well as all isospin-violating effects. Our expressions agree
with those of Ref.~\cite{deA} and Ref.~\cite{Ali98}.

%%%%%%%%%%%%%%%%%% REFERENCES %%%%%%%%%%%%%%%%%%%%%%%%%%%%

\newpage

\noindent {\bf FIGURES}

$\,$

\vspace{3cm}

\vspace{1cm}

\end{document}